\documentclass[final,5p,times,comma,round,numbers,twocolumn,sort&compress]{elsarticle}
\usepackage{amsmath}
 \pdfoutput=1
\usepackage{color}
\usepackage{caption}
\usepackage{colortbl}
\usepackage{hyperref}
\usepackage{ragged2e}
\usepackage{enumitem}
\usepackage{chapterbib}
\usepackage{cleveref}
\usepackage{tabularx}
\usepackage{longtable}
\usepackage{multirow}
\usepackage{xcolor}

\renewcommand\thesubsection{\thesection.\arabic{subsection}}
\usepackage{xpatch}
\usepackage[normalem]{ulem}
\xpatchcmd{\MaketitleBox}{\hrule\vskip12pt}{\vspace{-2\baselineskip}}{}{}
\xpatchcmd{\MaketitleBox}{\hrule}{}{}{}
\usepackage{array}
\newcolumntype{P}[1]{>{\centering\arraybackslash}p{#1}}
\usepackage{etoolbox}
\usepackage{comment}
\patchcmd{\abstract}{Abstract}{}{}{}

\usepackage{soul}

\renewcommand\thesection{\Roman{section}}
\journal{Nature Materials} 

\begin{document}
\renewcommand{\abstractname}{\vspace{-\baselineskip}}

\title{\vspace{-4em}\LARGE Measurement of competing pathways in a shock-induced \\phase transition in zirconium by 
	femtosecond diffraction}
\author{Saransh Singh\fnref{label1}\corref{cor}}
\ead{saransh1@llnl.gov}
\author{Martin G. Gorman\fnref{label1}}
\author{Patrick G. Heighway\fnref{label2}}
\author{Joel V. Bernier\fnref{label1}}
\author{\\ David McGonegle \fnref{label2,label3}}
\author{Hae Ja Lee\fnref{label4}}
\author{Bob Nagler\fnref{label4}}
\author{Jon H. Eggert\fnref{label1}}
\author{Raymond F. Smith\fnref{label1}}

\cortext[cor]{Corresponding author}
\fntext[label1]{Lawrence Livermore National Laboratory, Livermore, CA, USA}
\fntext[label2]{Department of Physics, Clarendon Laboratory, University of Oxford, Oxford, UK}
\fntext[label3]{Oxford Centre for High Energy Density Science, Department of Physics, Clarendon Laboratory, University of Oxford, Oxford, UK}
\fntext[label4]{Linac Coherent Light Source, SLAC National Accelerator Laboratory, Menlo Park, CA, USA}
\maketitle
\noindent
\textbf{The traditional picture of solid-solid phase transformations assumes an ordered parent phase transforms into an ordered daughter phase via a single unique pathway. Zirconium and its prototypical phase transition from hexagonal close-packed (hcp) to simple hexagonal (hex-3) structure has generated considerable controversy over several decades regarding which mechanism mediates the transformation. However, a lack of \emph{in situ} measurements over the relevant atomistic timescales has hindered our ability to identify the true pathway. In this study, we exploit femtosecond X-ray diffraction coupled with nanosecond laser compression to give unprecedented insights into the complexities of how materials transform at the lattice level. We observe single-crystal zirconium changing from hcp to a hex-3 structure via not one but three competing pathways simultaneously. Concurrently, we also observe a broad diffuse background underlying the sharp Bragg diffraction during the transition. We corroborate our observation of the diffuse signal with multimillion-atom molecular dynamics simulations using a machine-learned interatomic potential. Our study demonstrates that the traditional mechanistic view of transitions may fail for even an elemental metal and that the mechanisms by which materials transform are far more intricate than generally thought.
}

\smallskip
\noindent
Solid-solid phase transitions have been an active area of scientific inquiry for the last hundred years \cite{Bridgman1912,Bain1924,Kurdjumow1930,Shoji1931,Wassermann1933,Burgers1934,Nishiyama1934}. The lattice-level restructuring brought on by applying sufficient heat or pressure is conventionally measured using X-ray diffraction techniques, yielding direct interatomic separation and coordination measurements. Ultrabright sources such as X-ray free-electron lasers (XFELs) make it possible to realize such measurements \textit{in situ} in solids dynamically compressed to extreme, planetary-scale pressures over mere nanoseconds. Recent research efforts have focused increasingly on using diffraction to pin down not only the structure of exotic high-pressure phases but the atomistic pathway by which that phase is reached \cite{Yagi1992,Hawreliak2006,Pandolfi2022}. 

The conventional picture of the atomistic pathways is based upon atomic mapping from an ordered parent phase to an ordered daughter phase via lattice strain and atomic shuffles. Each transition pathway is differentiated by the orientation with which it leaves the daughter phase relative to its parent. Thus, by measuring the orientation relationship (OR) between specific crystallographic planes and directions within the parent and daughter phases using diffraction, one can experimentally constrain the atomistic transition mechanism.

Due to the commercial and technological importance of zirconium (Zr), titanium (Ti), and their alloys, the transformation mechanism between the ductile $\alpha-$phase (hcp) and the brittle high-pressure $\omega-$phase (simple hexagonal) has been the subject of intense theoretical and experimental scrutiny \cite{SILCOCK1958481,SARGENT1971,RABINKIN1981,UZ1973,Kutsar1990,Swinburne2016,Song1995,Jyoti1997,jyoti2008,osu1070481734,Trinkle2003,Wenk2013,ADACHI20151,wang2019,ZONG2014a,ZONG2014b,Guan2016,Rawat2017,ZONG2019}. Existing literature suggests that two distinct mechanisms can mediate the $\alpha\to\omega$ phase transition. The majority of shock-compression studies on recovered samples have observed what we will call the variant I OR \cite{Song1995,Jyoti1997,jyoti2008}. In contrast, the majority of static-compression and high-pressure torsion studies report variant II \cite{SILCOCK1958481,SARGENT1971,RABINKIN1981,Wenk2013,ADACHI20151,wang2019} (see supplementary materials Table~\ref{Table:OR}). However, several studies deviate from this trend \cite{UZ1973,Kutsar1990,Swinburne2016} for reasons that remain unclear. A complete understanding of this model phase transition thus remains elusive.

In this study, we approach the $\alpha\to\omega$ phase transition from a new direction by combining uniaxial laser-shock compression of high-purity, well-oriented single-crystal Zr with \emph{in-situ} diffraction measurements using a high-brightness femtosecond X-ray free electron laser (XFEL) source. We produce near-instantaneous snapshots of atoms in the high-pressure shocked state as the material is transforming using the femtosecond probe duration, revealing lattice-level details of the phase transition mechanism.
\begin{figure*}[t!]
	\includegraphics[width=1\textwidth]{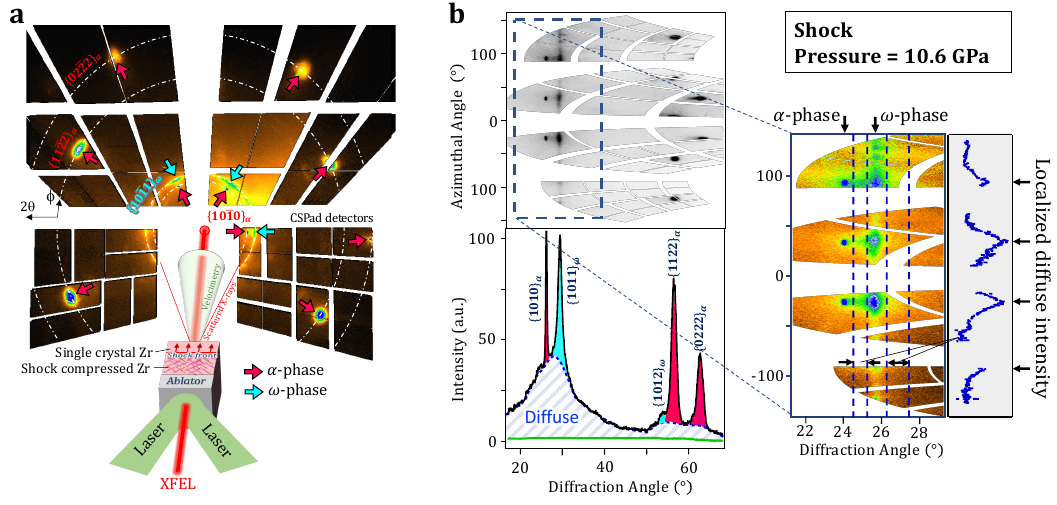}
	\caption{\label{fig:Fig_1} 
		\textbf{$\vert$ Experimental setup for X-ray diffraction measurements of shock-compressed single-crystal Zr samples.} 
		\textbf{a.} The 10 keV, 50 fs output of the LCLS-XFEL probed the shock-compressed Zr and produced an X-ray diffraction signal as recorded on CSPad detectors \cite{Hermann2013}.~Here, the diffraction angle ($2\theta$) increases radially, with azimuthal angle ($\phi$) defined around the Debye-Scherrer cones.~This data provides information on crystal structure, sample density, and microstructural texture, while velocimetry measurements provided information on shock timing and pressure uniformity during the X-ray probe period \cite{Celliers2004}.~The diffraction data shown [10.6 GPa] is consistent with multiple reflections from the compressed $\alpha$ phase (red arrows) and a single intense peak from the $\omega-$phase (cyan arrows). \textbf{b.} We present the same diffraction data in polar coordinates. The Debye-Scherrer rings appear as straight lines in this view. We show the $\phi$-averaged lineout in the lower panel, where the green curve represents the detector background. The peaks from the $\alpha$ and $\omega-$phase are shaded with red and cyan colors, respectively. We reproduce an enlarged view of the lower diffraction-angle region in the inset. The lineout to the right of the inset shows the azimuthal variation of the diffuse signal between the Bragg peaks noted by the dashed vertical lines.}
\end{figure*}
We performed laser-driven shock-compression experiments at the Matter in Extreme Conditions (MEC) endstation of the Linac Coherent Light Source (LCLS) \cite{Nagler2015,brown2017}. We compressed eight targets with peak pressures ranging from 5.3 to 21.3 GPa. We show a schematic of the experimental setup and target design used at MEC in Fig.~\ref{fig:Fig_1}\textbf{a}.~The target consisted of a plastic ablator and a 40-$\mu$m-thick Zr single crystal with the $[0001]$ direction parallel to the sample normal (shock direction). We used an ultrabright XFEL pulse incident normal to the target to perform volume-integrated X-ray diffraction. We used Cornell-Stanford Pixel Array Detectors (CSPADs) to record the scattered X-rays from the shocked Zr sample in transmission geometry \cite{Hermann2013}.

In Fig.~\ref{fig:Fig_1}\textbf{a}, we show the 2D orthographic view of diffraction data for a shock pressure of 10.6~GPa (marginally above the phase transition pressure). While the Zr sample loses its simple single crystalline nature upon shock-compression, it retains a highly oriented microstructure, as seen in the localized intensity around the Debye-Scherrer diffraction rings. At this pressure, we observe diffraction signals from both the compressed $\alpha$ and $\omega-$phases. We do not measure any diffraction from the uncompressed material ahead of the shock front due to the highly oriented nature of the zirconium foil (supplementary materials Fig.~\ref{fig:SM_CSPAD}). We also observe a diffuse background signal superimposed on top of sharp Bragg peaks. This diffuse signal (marked by dashed lines in the Fig.~\ref{fig:Fig_1}\textbf{b} inset) in the intervening regions between the Bragg peaks is observed to be non-uniform azimuthally and has an angular period of $60^{\circ}$ indicating a sixfold symmetry. The azimuthally localized diffuse signal reveals itself only in our single-crystal scattering configuration; we do not observe it in our polycrystalline Zr diffraction data (see supplementary materials Fig.~\ref{fig:Poly_Zr} and accompanying text).

\begin{figure*}[ht!]
	\centering
	\includegraphics[width=1\textwidth]{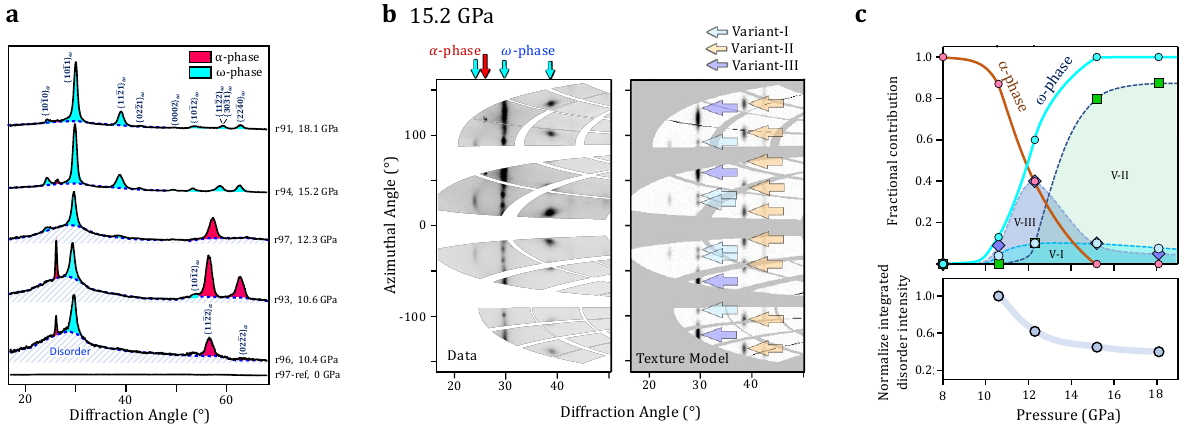}
	\caption{\label{fig:Fig_2}
		\textbf{$\vert$ Evolution of diffraction signal as a function of shock pressure.} \textbf{a.} Waterfall plot of azimuthally averaged lineouts from single crystal Zr targets. The diffuse scattering signal decreases with increasing pressure, with a corresponding increase in the number of peaks observed for the $\omega-$phase. \textbf{b.} Experimental and modeled diffraction patterns from shock-compressed single-crystal zirconium at 15.2~GPa. We modeled the crystallographic texture of the $\omega-$phase as the sum of unimodal distributions centered around the $12$ orientations of the variant I OR, $3$ orientations of variant II OR, and $6$ orientations of the variant III OR. We show the region with diffraction from the $\{10\bar{1}0\}_{\alpha}$, $\{10\bar{1}1\}_{\omega}$ and $\{11\bar{2}1\}_{\omega}$ planes. Contributions to the $\{10\bar{1}1\}_{\omega}$ diffraction from the ORs of variants I and III are azimuthally separated and indicated by the cyan and blue arrows, respectively. Diffraction from the $\{11\bar{2}1\}_{\omega}$ plane originates entirely from variant II OR. \textbf{c.} Volume fraction of each OR formed within the $\omega-$phase shows a strong pressure dependence. We observe the dominance of variant III OR at low pressures, which variant II OR eventually supplants at higher pressures.}
\end{figure*}

We show the azimuthally averaged lineouts at five pressures above the $\alpha-\omega$ phase boundary in Fig.~\ref{fig:Fig_2}\textbf{a}. We have shaded the peaks originating from the $\alpha-$phase in red and the $\omega-$phase in cyan. The $\alpha-$phase, which dominates the diffraction signal at 10.6 GPa, gradually recedes in intensity with increasing pressure. The diffraction signal at 18.1 GPa is almost exclusively from the high-pressure $\omega-$phase. We also observe a reduction in the integrated diffuse signal (hatched area) with an increase in pressure.

To accurately account for the observed intensity distribution in each Debye-Scherrer ring, we constructed a forward model to calculate the expected diffraction signal for our experimental geometry. This model accounts for X-ray beam energy and direction, detector positions, and crystallographic information about the sample, such as crystal structure, unit cell dimensions on compression, phase volume fraction, and crystallographic texture of each phase (see Methods section for details).
We present the results of the forward model calculation for the 15.2 GPa shot in Fig.~\ref{fig:Fig_2}\textbf{b}. We determined the phase fractions of the $\omega-$phases to be close to 100\% for this shot. The azimuthal intensity distribution is consistent with the presence of $\omega-$phase orientations from three distinct phase transition pathways: variants I, II, and III. For clarity, we only show the variant contributions to two reflections. While variants I and II have been reported in numerous studies, variant III has never been observed experimentally. We present the pressure dependence of the phase-fraction evolution, the relative abundance of the different phase transition ORs within the $\omega-$phase, and the integrated diffuse background signal in Fig.~\ref{fig:Fig_2}\textbf{c}. The phase fraction of the $\alpha-$phase decreases with increasing pressure as expected. In the $\omega-$phase volume, we observe three competing phase transition pathways vying for dominance. variant III dominates at lower pressures and is eventually superseded by variant II as the pressure increases. We refer the readers to supplementary materials Fig.~\ref{fig:phase_fraction} for our semi-quantitative method of determining the phase fractions of different phases and the relative abundance of the various phase transition ORs and Fig.~\ref{fig:Fig_2_old} for a comprehensive collection of the forward model results with increasing pressure.

\begin{figure*}[ht!]
	\centering
	\includegraphics[width=1\textwidth]{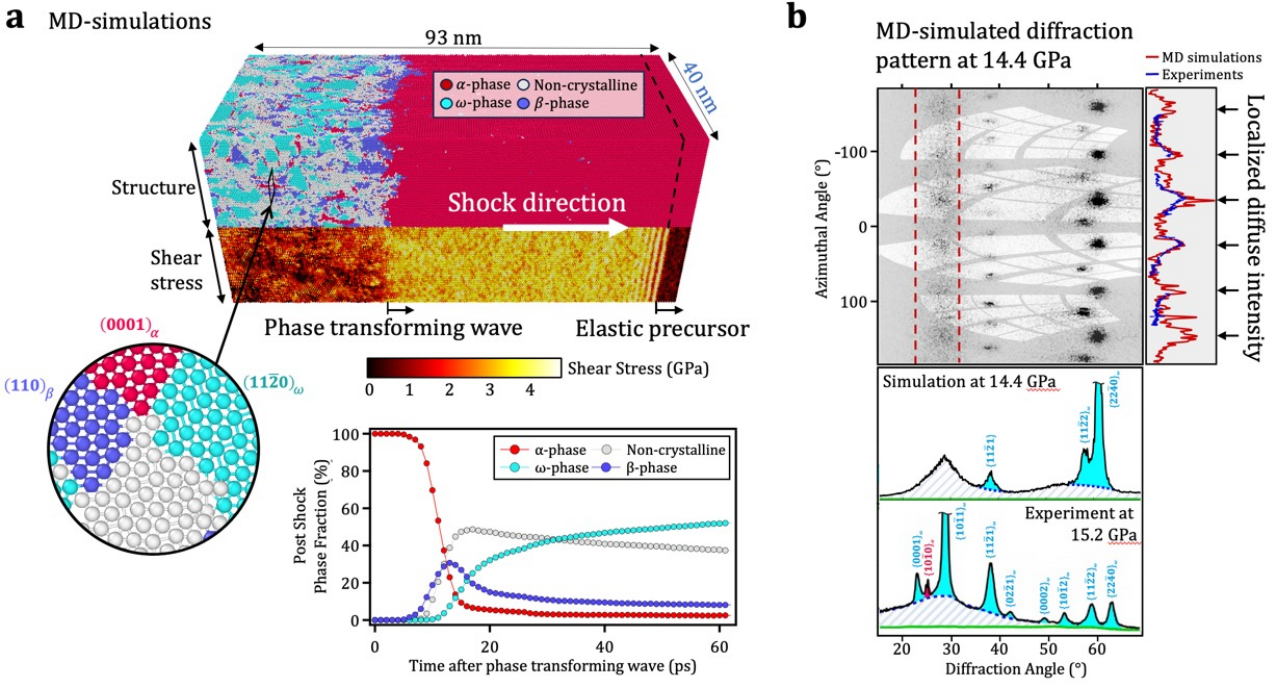}
	\caption{\label{fig:Fig_4}
		\textbf{$\vert$ Molecular dynamic simulations with machine learning potential.} \textbf{a.} Visualization of a Zr single crystal with initial dimensions of $40\times40\times100$~nm\textsuperscript{3} shock-compressed along [0001] to 14~GPa, with atoms colored according to their local phase (above) and shear stress (below) after 18~ps, performed using \textsc{ovito} \cite{OVITO}. Below is the time evolution of the $\alpha$, $\beta$, $\omega$, and non-crystalline phase fractions (the last group includes both partially disordered nano-clusters and atoms on grain boundaries). The circular inset shows the local arrangement of atoms in a region of phase coexistence, viewed in a plane normal to the shock. \textbf{b.} Synthetic diffraction generated 158~ps into the simulation using 10~keV X-rays. The top panel shows the full diffraction pattern in polar coordinates, with a semitransparent mask added to simulate gaps in the CSPad detectors. The lower panel compares azimuthally averaged lineouts from the simulation (14~GPa) and experiment (15.2~GPa). The right panel shows the azimuthal variation of the simulated diffuse signal centered at $2\theta\sim30^\circ$, with averages taken over the windowed region (red-boxed region). We overlay the azimuthal distribution of the azimuthally localized diffuse signal from the experimental data (shot 93) in blue.}
\end{figure*}

The azimuthally localized yet diffuse diffraction alludes to further complexity during the phase transition. To investigate the origin of this diffuse signal, we performed large-scale classical molecular dynamics (MD) simulations of [0001]-oriented Zr single crystals with initial dimensions of $40\times40\times100$~nm\textsuperscript{3} shocked to 14~GPa using \textsc{LAMMPS} \cite{LAMMPS}. We modeled the interatomic interactions using the machine-learned potential of Zong \textit{et al.}\ \cite{ZONG2019}, which was tailored to model allotropy in Zr at pressures of up to 30~GPa. These simulations allow us to see in microscopic detail the dynamics unfolding in the wake of the shock and to analyze the structure factor of the complex atomistic configuration that results.

We show in Fig.~\ref{fig:Fig_4}\textbf{a} a visualization of the computational cell 18~ps into the simulation, with atoms colored according to their local phase (above) and shear stress (below). The shock is led by an elastic precursor, behind which the pressure and shear stress are elevated to 10.0~GPa and 3.5~GPa, respectively, but the crystal retains a strained $\alpha-$phase structure. Trailing this precursor is a phase transition front that raises the pressure to 14~GPa while relaxing the shear stress to 0.9~GPa, heralding the onset of the $\alpha\to\omega$ transition. The parent $\alpha$ phase first transforms to a short-lived ($\sim$10~ps) intermediate $\beta$ (bcc) phase before taking one of two pathways: the $\beta$ nano-grains either promptly transform to the $\omega$ structure via a displacive mechanism~\cite{UZ1973}, or they rapidly amorphize and subsequently reconstitute into an anisotropic yet non-crystalline structure via slower diffusive motion. The resulting clusters of partially disordered atoms form 10\% of the cell by mass and occupy the interstices between the larger $\omega-$grains (which constitute 60\%). Though non-crystalline, the atomistic structure of these nano-clusters shows orientational order and a structure factor qualitatively similar to that of the $\omega-$phase (see supplementary materials Fig.~\ref{fig:pair-correlation} for the radial and angular distribution functions and Fig.~\ref{fig:structure_factors} for the structure factors of the $\omega$ and non-crystalline structures).

In Fig.~\ref{fig:Fig_4}\textbf{b}, we show synthetic diffraction generated by the phase-transformed region of the simulation cell with 10 keV X-rays. Sharp diffraction peaks attributable to $\omega$ grains with the variant II OR are visible at Bragg angles similar to those measured in the experiment and at identical azimuthal angles. The difference in peak intensities is due to the strong sensitivity of the Bragg condition to the $\omega$ phase's $c/a$ ratio, which at 14~GPa takes values of 0.605 and 0.621 in the experiment and simulation, respectively. In addition to these Bragg peaks, we observe a diffuse diffraction signal centered at $29^\circ$ akin to that seen in the data. Like the experimental signal, this diffuse diffraction is anisotropic and exhibits the same sixfold modulations around the Debye-Scherrer ring. While it is tempting to attribute the diffuse scattering to the partially disordered nano-clusters, the truth is subtler: both the crystalline $\omega-$phase and disordered material contribute to the diffuse signal.
Moreover, their contributions do not decompose additively, suggesting they diffract coherently. The signal in question is diffuse because the active scattering vectors are situated far from the sharp maxima in the aggregate's structure factor. Our simulations suggest that we are measuring tenuous interpeak scattering intensity that is dwarfed by intense Bragg diffraction in a conventional polycrystal experiment.

The new variant III OR we observe in our experiments has never been observed in quenched $\omega-$Zr captured from recovery experiments. This observation is in contrast to numerous experimental static and dynamic compression studies reporting variant I and variant II ORs \cite{SILCOCK1958481,SARGENT1971,UZ1973,RABINKIN1981,Kutsar1990,Song1995,Jyoti1997,jyoti2008,Wenk2013,ADACHI20151,wang2019}.~In a recovery study of shock and pressure released Zr, samples that completely transformed to the $\omega-$phase during shock loading retained between $0$ to $48\%$ of the high-pressure phase in recovered samples \cite{gorman2020}. The lack of any experimental observation of the variant III OR in recovered samples could be related to an OR dependence of the $\omega\rightarrow\alpha$ reverse phase transition. Upon pressure release, we hypothesize a complete reversion of the variant III $\omega-$phase orientations to the $\alpha-$phase, but hysteresis for the other variants.

In our MD simulations, the vast majority (99\%) of the $\omega-$phase assumes the variant II OR, with only trace amounts of variant I present. This prediction contrasts with the multi-mechanism picture painted by the experiment. However, the simulation properties can rationalize this behavior. Zong \emph{et al.} \cite{ZONG2019}, in their MD study using the identical interatomic potential as our study, have reported a switch in the $\omega-$phase OR from variant II to variant I upon artificially reducing shear stress in their simulations. We posit that variant I is made rare by the artificially high shear-stress environment created in these idealistic simulations in perfect, defect-free crystals. In the presence of defects, some plasticity before the phase transition will bring about the low shear-stress environment that favors variant I OR over variant II. Recent static high-pressure diffraction measurement by Singh \emph{et al.}\ \cite{Singh2023} on highly oriented Zr also supports this physical model. In contrast to other static-compression experiments \cite{SILCOCK1958481,SARGENT1971,RABINKIN1981,Wenk2013,ADACHI20151,wang2019}, the authors in that study observed variant I OR under static-compression when neon was used as a pressure medium to ensure purely hydrostatic conditions. We believe the absence of variant III in our simulations results from the interatomic potential's inability to accurately model the energy landscape of Zr along the variant III pathway. As this transformation pathway was undiscovered until now, its configuration space did not form part of the database on which the potential was trained. Exploration of this new pathway via first-principles atomistic simulations is warranted.

Our MD simulations predict a small amount of the intermediate bcc $\beta-$phase in the tens of \textit{ps} timescale. Our forward diffraction model predicts that the $\{220\}_{\beta}$ reflection should be observable in our experiments for the expected Burger's OR \cite{Burgers1934} between the hcp $\alpha-$phase and the bcc $\beta-$phase. Although not conclusive, our diffraction data hints at the presence of this intermediate bcc phase. We refer the readers to the supplementary materials Fig.~\ref{fig:s96}, \ref{fig:beta} and accompanying text for more details.

In summary, the dynamic behavior of the elemental metal Zr under shock compression is far more intricate than previously thought, involving multiple competing phase transition pathways and a partially disordered state in a single \textit{ns} shock volume. While we initially attempted to use the more conventional EAM potential for Zr, only the machine-learned interatomic potential partially reproduced our experimental observations. Our experiments and analysis highlight the importance of \textit{in situ} measurements on highly oriented crystals in validating our best theoretical models at extreme conditions. While our results focus on zirconium, we expect our results to apply to the transformation dynamics of other material systems at high pressure and temperature.

\vspace{-1em}
\noindent
\bibliographystyle{science}
\bibliography{references}
\medskip
\noindent

\noindent\bf Methods

\footnotesize{
	\noindent
	\textbf{Target.} \normalfont{A schematic of the experimental setup and target design used at MEC is shown in Fig. \ref{fig:Fig_1}.~The target consisted of a 50-$\mu$m thick polyimide ablator, a 40-$\mu$m-thick Zr [0001] single crystal.~The single crystal Zr samples were supplied by Princeton Scientific Corp. The samples were orientated to with $<$ 1$^{\circ}$ of the [0001] axis. An impurity analysis confirmed $>$99.95\% purity and an ambient-pressure density of 6.52 g/cm$^3$.\\
		
		\noindent
		\textbf{Forward model.} ~Given a material system with a known crystal structure, lattice parameters, temperature, and crystallographic orientation distribution function, the forward model computes the expected diffraction signal in $2\theta-\phi$ space. In any diffraction experiment in the transmission geometry and a single sample view, only a single ring of the complete pole figure is observable. Therefore, given an orientation distribution function, the forward model modulates the powder diffraction intensity in the azimuthal direction by the observable pole density for each one of the reflections. Fig.~\ref{fig:FM_alpha} shows the complete pole figures for a unimodal orientation distribution centered around the starting orientation of the $\alpha-$phase. We show similar pole figures for the $\omega-$phase for the three observed ORs in~\Cref{fig:FM_omega_var1,fig:FM_omega_var2,fig:FM_omega_var3}. The dotted magenta circle shows the observable ring of the complete pole figure for each of the reflections. The computation proceeds by the following steps:
		\begin{enumerate}[leftmargin=*,noitemsep,topsep=0pt]
			\item For a given material and reciprocal lattice reflection, compute the powder diffraction intensity. The factors include the volume fraction of the phase, unit cell volume, structure factor, Lorentz-polarization factor, the multiplicity of the reflection, and the Debye-Waller factor.
			\item Modulate the powder diffraction intensity in the azimuthal direction by multiplying the magnitude of the pole density around the observable ring.
			\item Plot the intensity from the previous step as a pseudo-voigt function in $2\theta$ dimension. The full-width half maxima is a function of $2\theta$ via the Cagliotti parameters \cite{CAGLIOTI1958} and particle size, micro-strain parameters. The Thomas-Cox-Hastings profile function \cite{Thompson1987} gives the mixing factor.
			\item Apply the mask in $2\theta-\phi$ derived from the composite CSPAD detector.
		\end{enumerate}
		We performed all computations relating to the orientation distribution function and pole density using the finite element representation of the Rodrigues space fundamental zone \cite{KUMAR1998,Barton2002}.\\
		
		\vspace{-0.5em}
		\noindent
		\textbf{Laser compression.} ~The output of the 527-nm drive laser at MEC, temporally shaped into a 15-ns flattop pulse, was focused into a 250-$\mu$m spot on the front of the polyimide. This generated an ablatively-driven shock into the target assembly, which an applied uniaxially load along the c-axis of the Zr crystal.~Shock pressure in the sample is directly proportional to the laser intensity, and, therefore precise control of the on-target energy allowed for precise tuning of the pressure within the Zr sample \cite{brown2017}. We set the X-ray probe time in our experiments to ensure the same shock-compressed Zr volume from shot-to-shot, as determined by the velocimetry measurements (Fig. \ref{fig:VISAR}a).\\
		
		\vspace{-0.5em}
		\noindent
		\textbf{Molecular dynamics simulations.} ~Large-scale classical molecular dynamics (MD) simulations of dynamically compressed zirconium were performed using the open-source code \textsc{LAMMPS}~\cite{LAMMPS}. Interatomic interactions were modeled using the machine-learned potential developed by Zong \textit{et al.}~\cite{ZONG2019}. This short-range classical potential is trained on a reference energy database generated using \textit{ab initio} MD simulations and nudged elastic band calculations. The database includes atomistic configurations representing the $\alpha$, $\beta$, and $\omega$ phases of Zr -- as well as those found along the high-symmetry transition pathways from $\beta\to\alpha$, $\beta\to\omega$, and $\alpha\to\omega$ -- at pressures of up to 30 GPa. In our estimation, this potential is best suited to treating allotropy in Zr at high pressures.
		
		The crystals we simulate are initially defect-free blocks of $\alpha$-Zr with their [0001] direction aligned with the compression axis, $z$. The crystals are subjected to periodic boundary conditions in the transverse directions, $x$ and $y$, to simulate laterally confining material present behind a uniform, uniaxial shock front. All crystals span at least 40 x 40 nm\textsuperscript{2} in the $x$ and $y$ directions to mitigate finite-size effects that might otherwise influence grain growth. Shock waves were driven into the crystal using a momentum mirror ramped to the final particle velocity $U_P$ over 250~ps. A 1-fs timestep is used throughout.
		
		
		To relax computational costs, we use an iterative scheme whereby the length of the crystal is grown incrementally at a rate that keeps pace with the shock front. At every timestep, there exists a tracer layer at a distance $\delta = U_S^e t_{\text{eq}}$ from the crystal's rear surface, where $U_S^e$ is the speed of the elastic precursor wave, and $t_{\text{eq}}$ is the time required for a pristine block of crystal to thermalize ($\sim1$~ps). When the tracer velocity exceeds $0.5U_P^e$, a new block of Zr of thickness $\delta$ is seamlessly appended to the rear surface; the new material thermalizes for at least time $t_{\text{eq}}$ before the shock front sweeps through it. By comparing simulations carried out under an EAM potential with and without this incremental scheme, we have verified that the stress waves generated by the addition of new material (which cause oscillations in the pressure and shear stress of $\pm1\%$ and $\pm5\%$ of their peak values, respectively) do not substantively change the final thermodynamic state reached by the system. By building the `runway' required by the shockwave iteratively -- rather than constructing the crystal with its ultimate dimensions at the outset -- we avoid needless simulation of ambient material waiting to be compressed and thus double our computational efficiency.
		
		We also set a `cutoff time' after which we isolate a relatively small portion of the crystal far behind the shock front for study, and delete the remainder. Once a sufficiently large volume of phase-transformed Zr has accumulated near the front surface, we find the greatest $z$-coordinate at which the average particle velocity is no less than 95\% of the Hugoniot particle velocity $U_P$, insert at this limit a second momentum mirror co-moving with the first at fixed velocity $U_P$, and remove all atoms beyond the second mirror. The remaining block of Zr, which comprises at least 2 million atoms, evolves under constant-volume boundary conditions until the global phase fractions stabilize.
		
		To partition the crystal into its constituent $\alpha$, $\beta$, $\omega$, and non-crystalline phase fractions, we use the following pipeline [executed in \textsc{OVITO}~\cite{OVITO}]:
		\begin{enumerate}[leftmargin=*,noitemsep,topsep=0pt]
			\item Following Zong \textit{et al.}~\cite{ZONG2019,Zong2020}, we first classify the atoms according to their Ackland-Jones parameter (AJP) \cite{AcklandJones2006}. Atoms in the $\omega$ phase take AJP values of 2 or 3, depending on whether they are situated on $\omega$'s graphene-like planes or the basal planes that sandwich them, respectively. For brevity, we refer to atoms classified thus as `graphitic' and `basal,' respectively.
			\item Atoms in the $\omega$ phase are identified by analyzing their nearest and next-nearest neighbors' AJP values. We count the number of graphitic and basal atoms ($n_g$ and $n_b$, respectively) with which each central atom is coordinated within a cutoff radius of 3.5~\AA. Within this radius, basal atoms should have two basal neighbors and 12 graphitic neighbors; graphitic atoms should have five graphitic neighbors and six basal neighbors. Basal and graphitic atoms whose coordination numbers $(n_g,n_b)$ take the values $(2,12)$ or $(6,5)$, respectively, are assigned to the $\omega$ phase. We give a tolerance of $\pm1$ when comparing coordination numbers to combat thermal fluctuations.
			\item Atoms not assigned to the $\omega$ phase are subsequently classified by adaptive common-neighbor analysis (aCNA). Atoms are assigned to $\alpha$ (hcp), $\beta$ (bcc), or `other' structures. The third classification includes atoms in the bulk partially disordered phase, those found on grain boundaries, and `false negatives' whose local crystalline environment was so strongly perturbed by thermal fluctuations as to be unrecognizable to the AJP or aCNA.
			\item Atoms whose assigned phase differs from that of all but no more than one of its nearest neighbors are reassigned to the structure of those neighbors. This corrective step further mitigates misidentifications from thermal noise.
		\end{enumerate}
		Whenever identifying phases, we use atomic coordinates that have been time-averaged over 200~fs (approximately one oscillation period) to further reduce the number of false negatives incorrectly assigned a non-crystalline structure.
		
		To synthesize diffraction patterns from a set of (identical) atoms, we first calculate the ionic structure factor on the locus of $q$-vectors forming the Ewald sphere using the standard expression
		\begin{equation}
			S_i(\textbf{q}) = \sum_{\alpha=1}^N e^{-i \mathbf{q} \cdot \mathbf{r}_\alpha}\ ,
		\end{equation}
		where the $\{\mathbf{r}_\alpha\}$ are the (instantaneous) coordinates of the $N$ atoms of interest. The raw ionic scattering intensity $\lVert S_i(\mathbf{q})\rVert^2$ is calculated and post-multiplied by atomic-form, Lorentz-polarization, and self-attenuation factors.
		
	}
}

\noindent\textbf{Data availability}

\noindent
\footnotesize{\normalfont{The data supporting the findings of this study are available from the corresponding authors upon reasonable request.
	}
}

\noindent\textbf{Acknowledgements}

\noindent
\footnotesize{\normalfont{We thank the Linac Coherent Light Source (LCLS) operations staff and the Target  Engineering  Team at  Lawrence  Livermore  National  Laboratory  (LLNL)  for assistance in these experiments. The authors would also like to acknowledge Hongxiang Zong and Graeme J. Ackland for generously sharing their machine learning-derived interatomic potential for Zr and Justin Wark for his helpful suggestions during the writing of this manuscript. The research was supported by the Laboratory Directed Research and Development Program at LLNL (project nos.\ 17-ERD-014 and 21-ERD-032). This work was performed under the auspices of the US Department of Energy by Lawrence Livermore National Laboratory under Contract No.\ DE-AC52-07NA27344 (LLNL-JRNL-XXXXXX).} P.\ G.\ H.\ gratefully acknowledges the support of AWE via the Oxford Centre for High Energy Density Science (OxCHEDS).}

\noindent\textbf{Author Contributions}

\noindent
\footnotesize{\normalfont{S.S., M.G., P.G.H., D.M., J.B., R.B., J.E., and R.S. were involved in all aspects of the experimental design and analysis. B.N. and H-J.L.  were involved in all aspects of the experimental acquisition of data. The manuscript was written by S.S., M.G., P.G.H., J.E., and R.S.and reviewed by all authors.}}

\noindent\textbf{Competing Interests}

\noindent
\footnotesize{\normalfont{
		\noindent The authors declare no competing interests.
	}
}

\clearpage
\onecolumn
\noindent
\centering
\normalfont
\Large{\textbf{Measurement of competing pathways in a shock-induced \\phase transition in zirconium by 
		femtosecond diffraction}}
\medskip

\noindent
\normalsize{Saransh Singh, Martin G. Gorman, Patrick G. Heighway, Joel V. Bernier,\\ David McGonegle, Hae-Ja Lee, Bob Nagler, Jon H. Eggert, Raymond F. Smith}

\justifying
\bigskip
\noindent
\Large{Supplementary Materials}

\renewcommand{\thefigure}{S\arabic{figure}}

\renewcommand{\thefigure}{S\arabic{figure}}
\renewcommand{\thetable}{S\arabic{table}}
\renewcommand\thesection{S\arabic{section}}
\renewcommand\thesubsection{S\thesection.\arabic{subsection}}

\captionsetup[table]{name=Table,labelsep=space}
\captionsetup[Table]{labelfont=bf}

\setcounter{table}{0}
\setcounter{figure}{0}
\setcounter{table}{0}
\setcounter{section}{0}
\normalsize 

\medskip
\noindent
This Supplementary Material contains additional information relevant to the paper. We present a more complete set of experimental data collected and analysis conducted as part of this experiment. The document also contains all the relevant pole figure information to reproduce Fig.~\ref{fig:Fig_2}\textbf{b}, as well as the forward model results for the other shots. We also present our data hinting at the presence of a \textit{ns} persisting body-centered cubic (bcc) intermediate phase. \\


\noindent
\textbf{Supplementary Material Tables:}

\medskip
\noindent
\textbf{Table \ref{Table:OR}:} Summary of ORs reported for $\alpha\rightarrow\omega$ for Zr, Ti and their alloys.

\medskip
\noindent
\textbf{Table \ref{Table:OR_bcc}:} Summary of ORs reported for hcp $\rightarrow$ bcc.

\medskip
\noindent
\textbf{Table \ref{Table:lp}:} Lattice parameters for all data points collected in this study.

\medskip
\noindent
\textbf{Table \ref{Table:numvariants1}:} Number of orientation variants for $\alpha\rightarrow\omega$ phase transition ORs.

\medskip
\noindent
\textbf{Table \ref{Table:numvariants2}:} Number of orientation variants for $\alpha\rightarrow\beta$ phase transition ORs.

\bigskip
\noindent
\textbf{Supplementary Material Figures:}


\medskip
\noindent
\textbf{Figure \ref{fig:SM_CSPAD}:} 2-D diffraction signal and azimuthally averaged lineouts for the ambient single crystal zirconium.

\medskip
\noindent
\textbf{Figure \ref{fig:Poly_Zr}:} Diffraction lineouts for shock compression of polycrystalline-Zr samples showing no diffuse background.

\medskip
\noindent
\textbf{Figure \ref{fig:phase_fraction}:} Contribution of diffraction signal form $\alpha$ and different orientations of $\omega-$phase observed in our experiments.

\medskip
\noindent
\textbf{Figure \ref{fig:Fig_2_old}:} Simulated diffraction pattern using the forward model for different pressures.

\medskip
\noindent
\textbf{Figure \ref{fig:pair-correlation}:} Pair-correlation functions of simulated $\beta-$phase, $\omega-$phase, and partially disordered nanoclusters at 14~GPa.

\medskip
\noindent
\textbf{Figure \ref{fig:structure_factors}:} Comparison of the simulated structure factors in a high-symmetry plane for the $\omega$ nanocrystal and partially disordered nanoclusters.

\medskip
\noindent
\textbf{Figure \ref{fig:s96}:} A repeat shot for the the data presented in Figs.~\ref{fig:Fig_1}, \ref{fig:Fig_2_old}\textbf{a}.

\medskip
\noindent
\textbf{Figure \ref{fig:beta}:} Simulated pole figure and diffraction signal if an intermediate bcc $\beta-$phase was observed. 

\medskip
\noindent
\textbf{Figure \ref{fig:FM_alpha}:} Simulated pole figure for single crystal $\alpha-$Zr shock compressed to 10.6 GPa.

\medskip
\noindent
\textbf{Figure \ref{fig:FM_omega_var1}:} Simulated pole figures for high pressure $\omega-$Zr formed at 15.2 GPa for the variant I ORs observed in this study.

\medskip
\noindent
\textbf{Figure \ref{fig:FM_omega_var2}:} Simulated pole figures for high pressure $\omega-$Zr formed at 15.2 GPa for the variant II ORs observed in this study.

\medskip
\noindent
\textbf{Figure \ref{fig:FM_omega_var3}:} Simulated pole figures for high pressure $\omega-$Zr formed at 15.2 GPa for the variant III ORs observed in this study.

\medskip
\noindent
\textbf{Figure \ref{fig:VISAR}:} Representative VISAR trace and hugoniot points measured in our experiments.

\medskip
\noindent
\textbf{Figure \ref{fig:SM_PFs}:} Distribution of a few low index crystallographic plane normals for different proposed ORs for the $\alpha\rightarrow\omega$ phase transition in the literature. 

\begin{figure}
\centering
\includegraphics[width=1\textwidth]{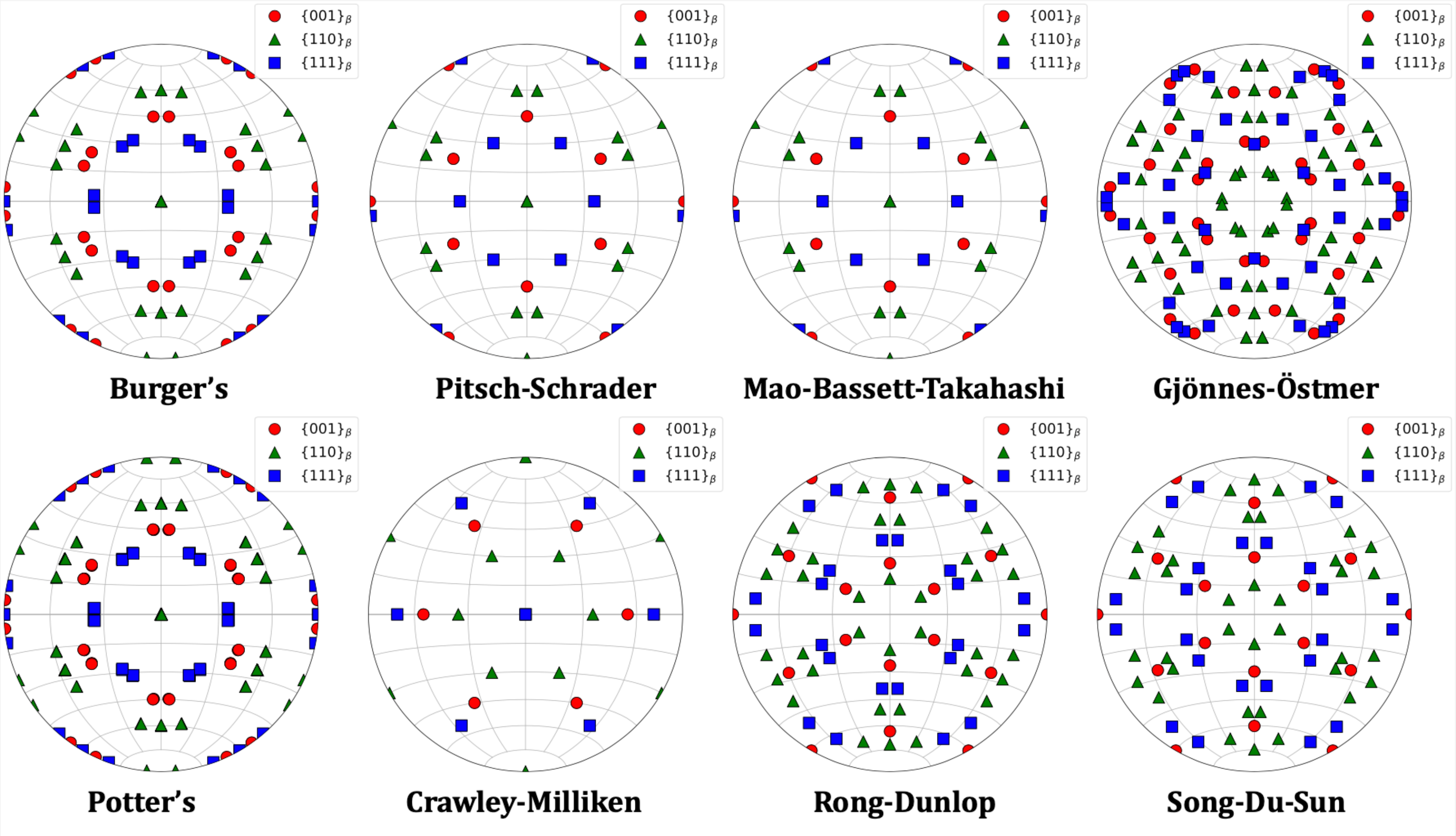}
\caption{Distribution of the $\{001\}, \{110\}$ and $\{111\}$ planes of the bcc $\beta-$phase for different ORs reported in experimental studies. References for each of these studies can be found in Table~\ref{Table:OR_bcc}. The Pitsch-Schrader \cite{Pitsch1958} is equivalent to Mao-Bassett-Takahashi OR \cite{Mao1967}. Direct comparison with observed diffraction can be used to identify the presence of an intermediate 
	$\beta-$phase lasting in the \textit{ns} timescale. \label{fig:SM_PFs_bcc}}
	\end{figure}

\medskip
\noindent
\textbf{Figure \ref{fig:SM_PFs_bcc}:} Distribution of a few low index crystallographic plane normals for different proposed ORs for the $\alpha\rightarrow\beta$ phase transition in the literature. 

\begin{figure}[!ht]
\centering
\includegraphics[width=0.5\textwidth]{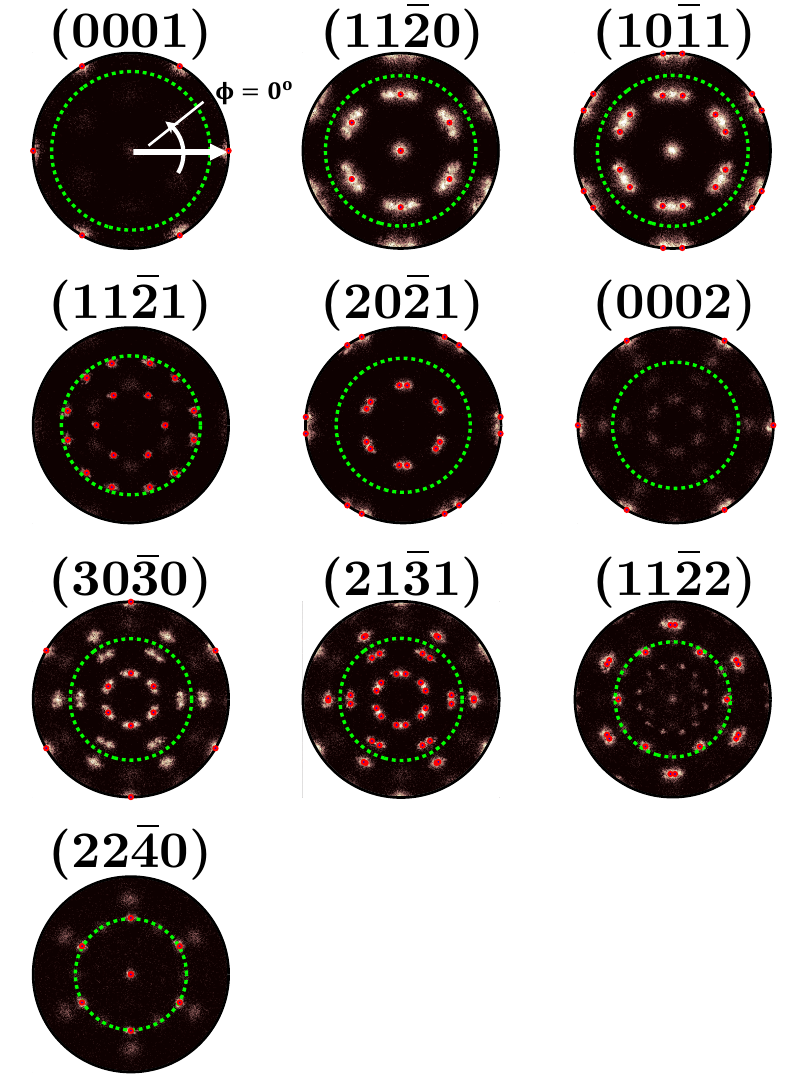}
\caption{Pole figures (linear scale) calculated from the structure factor of the 50 largest $\omega$ grains in the molecular dynamics simulation described in Fig.\ \ref{fig:Fig_4}. The observable rings in our experimental geometry for each reflection is shown by the dotted green circle. When the circle overlaps with non-zero intensity in the pole figure, then diffraction is observed for that reflection. The variation of intensity around the circle is reflected in the observed diffraction signal. Red circles mark the locations where the structure factor is expected to be strongest for variant II OR. Note that the $c/a$ ratio in the simulations is such that $\lvert\{11\bar{2}0\}\rvert\approx\lvert\{10\bar{1}1\}\rvert$ and $\lvert\{30\bar{3}0\}\rvert\approx\lvert\{21\bar{3}1\}\rvert$, making their plot plots practically inseparable.
	\label{fig:MD_pole}
}
\end{figure}

\medskip
\noindent
\textbf{Figure \ref{fig:MD_pole}:} Simulated pole figures for high pressure $\omega-$Zr for MD simulations in Fig. \ref{fig:Fig_2}.

\medskip
\noindent
\textbf{Figure \ref{fig:TAO}:} Supercell mapping using atomic shuffle and strain for the TAO-I $\alpha\rightarrow\omega$ phase transition mechanism giving rise to variant I OR.

\medskip
\noindent
\textbf{Figure \ref{fig:silcock}:} Supercell mapping using atomic shuffle and strain for the Silcock $\alpha\rightarrow\omega$ phase transition mechanism giving rise to variant II OR.

\newpage
\section*{\label{sec:OR}Orientation Relationships}
\begin{longtable}{|p{0.27\textwidth}|p{0.4\textwidth}| p{0.27\textwidth}|}
	\caption{\textbf{$\vert$ Overview of studies determining the OR during $\alpha\rightarrow\omega$ transformation in Ti, Zr and their alloys}\label{Table:OR}} \\
	\hline
	\centering
	\textbf{Authors} & \textbf{Experimental or Simulation details} & $\pmb{(\textrm{hkl})_{\alpha}}~~~~~ \vert\vert~~~~ \pmb{(\textrm{hkl})_{\omega}} \newline \pmb{[\textrm{uvw}]_{\ \alpha}}~~ \vert\vert~~~~ \pmb{[\textrm{uvw}]_{ \omega}}$ \\
	\hline
	Silcock \cite{SILCOCK1958481}\newline (1958) &  Experimental x-ray diffraction study of the Ti alloys. OR between $\alpha-$ and $\omega-$phase was theoretically determined by studying projections of different planes. & $(0001)_{\alpha} \vert\vert (11\bar{2}0)_{\omega}$ \newline $[11\bar{2}0]_{\alpha} \vert\vert [0001]_{\omega}$ \newline (\textbf{variant II}) \\
	\hline
	Sargent and Conrad \cite{SARGENT1971} \newline (1971) & Experimental study on pressure soaked at 2.8 GPa using Zone Axis diffraction patterns (ZADP) in a Transmission Electron Microscopy (TEM) & $(0001)_{\alpha} \vert\vert (11\bar{2}0)_{\omega}$ \newline (\textbf{variant II}) \\
	\hline
	Usikov and Zilbershtei \cite{UZ1973} \newline (1973) &  Experimental study on pressure soaked Ti and Zr using ZADPs in a TEM & $(0001)_{\alpha} \vert\vert (01\bar{1}1)_{\omega}$ \newline $[11\bar{2}0]_{\alpha} \vert\vert [10\bar{1}1]_{\omega}$ \newline (\textbf{variant I}) \\
	\hline
	Rabinkin, Talianker and Botstein \cite{RABINKIN1981} \newline (1981) & Static compression of Zr foils in a Diamond Anvil Cell (DAC). Orientation relation derived from selected area diffraction patterns of recovered samples & $(0001)_{\alpha} \vert\vert (11\bar{2}0)_{\omega}$ \newline $[11\bar{2}0]_{\alpha} \vert\vert [0001]_{\omega}$ \newline (\textbf{variant II}) \\
	\hline
	Kutsar et al. \cite{Kutsar1990} \newline (1990)   & Experimental shock compression using gas gun apparatus of Zr foils. Orientation relationship derived from ZADP in a TEM of recovered sample & $(0001)_{\alpha} \vert\vert (11\bar{2}0)_{\omega}$ \newline $[11\bar{2}0]_{\alpha} \vert\vert [0001]_{\omega}$ \newline (\textbf{variant II})  \\
	\hline
	Song and Gray \cite{Song1995}  \newline (1995) & Experimental shock compression using gas gun apparatus of Zr foils. Orientation relationship derived from ZADP in a TEM of recovered sample & $(0001)_{\alpha} \vert\vert (01\bar{1}1)_{\omega}$ \newline $[1\bar{1}00]_{\alpha} \vert\vert [11\bar{2}\bar{3}]_{\omega}$ \newline (\textbf{variant I}) \\
	\hline
	Jyoti et al. \cite{Jyoti1997}  \newline (1997) & Experimental shock compression using gas gun apparatus of Zr foils. Orientation relationship derived from ZADP in a TEM of recovered samples & $(0001)_{\alpha} \vert\vert (01\bar{1}1)_{\omega}$ \newline $[11\bar{2}0]_{\alpha} \vert\vert [10\bar{1}1]_{\omega}$ \newline (\textbf{variant I})  \\
	\hline
	Trinkle, Hennig and Wilkins \cite{Trinkle2003}  \newline (2003) & Geometric mapping of $\alpha$ and $\omega-$supercells together with \textit{ab initio} simulations & $(0001)_{\alpha} \vert\vert (01\bar{1}1)_{\omega}$ \newline $[11\bar{2}0]_{\alpha} \vert\vert [10\bar{1}1]_{\omega}$ \newline (\textbf{variant I} for lowest energy TAO-I pathway) \\
	\hline
	Jyoti et al. \cite{jyoti2008}  \newline (2008) & Experimental shock compression using gas gun apparatus of Zr foils. Orientation relationship derived from ZADP in a TEM of recovered samples & $(0001)_{\alpha} \vert\vert (01\bar{1}1)_{\omega}$ \newline $[11\bar{2}0]_{\alpha} \vert\vert [10\bar{1}1]_{\omega}$ \newline (\textbf{variant I}) \\
	\hline
	Wenk et al. \cite{Wenk2013}  \newline (2013) & Static compression of Zr wires using Deformation-DIA (D-DIA) and Diamond Anvil Cell (DAC) & $(0001)_{\alpha} \vert\vert (11\bar{2}0)_{\omega}$ \newline $[11\bar{2}0]_{\alpha} \vert\vert [0001]_{\omega}$ \newline (\textbf{variant II}) \\
	\hline
	Zong et. al. \cite{ZONG2014a}  \newline (2014) & Molecular dynamics of different single crystal Ti using a modified embedded atom method (MEAM) potential & $(0001)_{\alpha} \vert\vert (1\bar{1}00)_{\omega}$ \newline $[10\bar{1}0]_{\alpha} \vert\vert [11\bar{2}3]_{\omega}$ \newline (for c-oriented Ti; closely related to \textbf{variant III}) \newline $(0001)_{\alpha} \vert\vert (11\bar{2}0)_{\omega}$ \newline $[11\bar{2}0]_{\alpha} \vert\vert [0001]_{\omega}$ \newline (for $a$ and $a+b$ oriented Ti; \textbf{variant II}) \\
	\hline
	Zong et al. \cite{ZONG2014b}  \newline (2014) & Combination of experimental and simulations. Experimental shock compressed Ti samples using gas gun apparatus with orientation relationship for the reverse $\omega \rightarrow \alpha$ measured using synchrotron x-ray diffraction of recovered samples. Simulations performed using molecular dynamics simulations and a spline based MEAM potential for Ti & $(0001)_{\alpha} \vert\vert (11\bar{2}0)_{\omega}$ \newline $[11\bar{2}0]_{\alpha} \vert\vert [0001]_{\omega}$ \newline (\textbf{variant II}) \\
	\hline
	Adachi et al. \cite{ADACHI20151} \newline (2015) & Experimental study of $\alpha\rightarrow\omega$ phase transition during high-pressure torsion using Neutron diffraction & $(0001)_{\alpha} \vert\vert (11\bar{2}0)_{\omega}$ \newline $[11\bar{2}0]_{\alpha} \vert\vert [0001]_{\omega}$ \newline (\textbf{variant II}) \\
	\hline
	Guan and Liu \cite{Guan2016}  \newline (2016) & First-principles density functional theory calculation, including the heterojunction energy of the $\alpha$ and $\omega-$phases & $(11\bar{2}2)_{\alpha} \vert\vert (1\bar{1}00)_{\omega}$ \newline $[1\bar{1}00]_{\alpha} \vert\vert [11\bar{2}0]_{\omega}$ \newline (at higher pressures; closely related to \textbf{variant I})
	\newline $(10\bar{1}0)_{\alpha}\vert\vert(1\bar{1}00)_{\omega}$ \newline $[0001]_{\alpha}\vert\vert[11\bar{2}0]_{\omega}$ \newline (at lower pressures; \textbf{variant II}) \newline
	$(0001)_{\alpha}\vert\vert(02\bar{2}1)_{\omega}$ \newline
	$[11\bar{2}0]_{\alpha} \vert\vert [2\bar{1}\bar{1}0]_{\omega}$ \newline (via an intermediate cubic FCC phase)\\
	\hline
	Swinburne et al. \cite{Swinburne2016}  \newline (2016) & Experimental laser shock compression of rolled Zr foils coupled with \textit{in-situ} x-ray diffraction & $(10\bar{1}0)_{\alpha}~\vert\vert~ (10\bar{1}1)_{\omega} \newline [0001]_{\alpha}~ \vert\vert~ [1\bar{2}10]_{\omega}$ \\
	\hline
	Rawat and Mitra \cite{Rawat2017}  \newline (2017) & Molecular dynamics simulation of $c-$oriented Ti using $4$ different potentials & $(0001)_{\alpha} \vert\vert (01\bar{1}1)_{\omega}$ \newline $[11\bar{2}0]_{\alpha} \vert\vert [10\bar{1}1]_{\omega}$ \newline (\textbf{variant I}) \\
	\hline
	Zong et al. \cite{ZONG2019}  \newline (2019) & Molecular dynamics simulation of shock compression in differently oriented single crystal Zr using machine learned \textit{ab initio} potential & $(0001)_{\alpha}~\vert\vert~(11\bar{2}0)_{\omega}$ \newline (\textbf{variant II}) \\
	\hline
	Wang et al. \cite{wang2019} \newline (2019) & Experimental study of $\alpha\rightarrow\omega$ phase transition during high-pressure torsion using high energy x-ray diffraction & $(0001)_{\alpha} \vert\vert (11\bar{2}0)_{\omega}$ \newline $[11\bar{2}0]_{\alpha} \vert\vert [0001]_{\omega}$ \newline (\textbf{variant II}) \\
	\hline
	This study \newline (2023) & Experimental study of $\alpha\rightarrow\omega$ phase transition using laser compression and \textit{in-situ} X-ray diffraction on [0001] oriented Zr single crystals & $(0001)_{\alpha} \vert\vert (1\bar{1}00)_{\omega}$ \newline $[11\bar{2}0]_{\alpha} \vert\vert [11\bar{2}3]_{\omega}$ \newline (\textbf{variant III}) \\
	\hline
\end{longtable}

\begin{longtable}{|p{0.27\textwidth}|p{0.4\textwidth}| p{0.27\textwidth}|}
\caption{\textbf{$\vert$ Overview of studies determining the OR during hcp $\rightarrow$ bcc transformation}\label{Table:OR_bcc}} \\
\hline
\centering
\textbf{Authors} & \textbf{Experimental details} & $\pmb{(\textrm{hkl})_{\text{\small bcc}}}~~~~\vert\vert~~~~ \pmb{(\textrm{hkl})_{\text{\small hcp}}} \newline \pmb{[\textrm{uvw}]_{\text{\small bcc}}}~~ \vert\vert~~~~ \pmb{[\textrm{uvw}]_{\text{\small hcp}}}$ \\
\hline
Burger's \cite{Burgers1934}\newline (1934) &  Experimental X-ray diffraction study on the orientation relationship between the hcp $\alpha-$ and bcc $\beta-$phase in Zr. & $(110)_{\text{\small bcc}} \vert\vert (0001)_{\text{\small hcp}}$ \newline $[1\bar{1}1]_{\text{\small bcc}} \vert\vert [11\bar{2}0]_{\text{\small hcp}}$ \\
\hline
Pitsch-Schrader \cite{Pitsch1958}\newline (1958) & Experimental $\epsilon-$Fe & $(110)_{\text{\small bcc}} \vert\vert (0001)_{\text{\small hcp}}$ \newline $[1\bar{1}0]_{\text{\small bcc}} \vert\vert [10\bar{1}0]_{\text{\small hcp}}$ \\
\hline
Mao-Bassett-Takahashi \cite{Mao1967} \newline (1967) & Experimental X-ray diffraction study on the orientation relationship between the ambient $\alpha-$phase and high-pressure $\epsilon-$phase in Fe & $(110)_{\text{\small bcc}}  \vert\vert (0001)_{\text{\small hcp}}$ \newline 
$[001]_{\text{\small bcc}} \vert\vert [2\bar{1}\bar{1}0]_{\text{\small hcp}}$ \\
\hline
Gjönnes-Östmer \cite{Gjonnes1970}\newline (1970) & Experiment study of precipitates in Mg alloy & $(110)_{\text{\small bcc}}  \vert\vert (0001)_{\text{\small hcp}}$ \newline 
$[1\bar{1}2]_{\text{\small bcc}} \vert\vert [1\bar{2}10]_{\text{\small hcp}}$ \\
\hline
Potter \cite{potter1973} \newline (1973) &  Experimental study on orientation relationship between bcc $\alpha-$phase in Vanadium-Nitrogen system and hcp V$_{3}$N precipitate using transmission electron microscopy & $(110)_{\text{\small bcc}}  \vert\vert (1\bar{1}01)_{\text{\small hcp}}$ \newline $[1\bar{1}1]_{\text{\small bcc}} \vert\vert ~~[11\bar{2}0]_{\text{\small hcp}}$ \\
\hline
Crawley-Milliken \cite{Crawley1974}\newline (1974) & Experimental study on Mg-alloy using selected area diffraction in the transmission eletron microscope & $(111)_{\text{\small bcc}}  \vert\vert (0001)_{\text{\small hcp}}$ \newline $[11\bar{2}]_{\text{\small bcc}} \vert\vert ~~[2\bar{1}\bar{1}0]_{\text{\small hcp}}$ \\
\hline
Rong-Dunlop \cite{rong1984} \newline (1984) &  Experimental study on orientation relationship between bcc ferrite and hcp M$_{2}$C (M = Cr, Mo, Fe) precipitates in ASP23 high strength steel using transmission electron microscopy & $(021)_{\text{\small bcc}} \vert\vert (0001)_{\text{\small hcp}}$ \newline $[100]_{\text{\small bcc}}\vert\vert [11\bar{2}0]_{\text{\small hcp}}$ \\
\hline
Song-Du-Sun \cite{Song2002} \newline (2002) & Experimental study on orientation relationship between hcp M$_{2}$C (M = Mo, V) carbide precipitates and bcc ferrite matrix in M50NiL low carbon bearing steel using transmission electron microscopy & $(010)_{\text{\small bcc}} \vert\vert (01\bar{1}1)_{\text{\small hcp}}$ \newline $[100]_{\text{\small bcc}} \vert\vert [2\bar{1}\bar{1}0]_{\text{\small hcp}}$ \\
\hline
\end{longtable}

\section*{\label{sec:lp}Lattice Parameters}
\noindent
\begin{center}
\begin{longtable}{
	|P{0.08\textwidth}|P{0.075\textwidth}|P{0.35\textwidth}|P{0.35\textwidth}|}
\caption{\textbf{Table of Lattice Parameters}\label{Table:lp}} \\
\hline
\textbf{Shot $\#$} &
\textbf{Pressure (GPa)} & 
\textbf{$\pmb{\alpha}-$phase lattice constants (\AA), density (g cm$^{-3}$), c/a ratio } &
\textbf{$\pmb{\omega}-$phase lattice constants (\AA), density (g cm$^{-3}$), c/a ratio} \\
\hline Shot 244 & 
5.3 &
$a = 3.17, c = 5.075$, $\rho$ = 6.86, c/a = 1.60 &
\textemdash \\
\hline

Shot 96 & 
10.4 &
$a = 3.17, c = 4.8$, $\rho$ = 7.25, c/a = 1.51 &
$a = 4.9, c = 2.96$, $\rho$ = 7.48, c/a = 0.604 \\
\hline

Shot 93 & 
10.6 &
$a = 3.168, c = 4.8$, $\rho$ = 7.26, c/a = 1.51 & 
$a = 4.9, c = 2.98$, $\rho$ = 7.34, c/a = 0.608 \\
\hline

Shot 97 & 
12.3 &
$a = 3.11, c = 4.8$, $\rho$ = 7.53, c/a = 1.54 & 
$a = 4.87, c = 2.95$, $\rho$ = 7.5, c/a = 0.606\\
\hline

Shot 92 & 
14.1 &
$a = 3.15, c = 4.8$, $\rho$ = 7.34, c/a = 1.52& 
$a = 4.86, c = 2.94$, $\rho$ = 7.56, c/a = 0.605\\
\hline

Shot 94 & 
15.2 &
\textemdash &
$a = 4.8, c = 2.96$, $\rho$ = 7.69, c/a = 0.617\\
\hline

Shot 91 &
18.1 &
\textemdash &
$a = 4.77, c = 2.96$, $\rho$ = 7.79, c/a = 0.621\\
\hline

Shot 99 & 
21.3 &
$a = 3.15, c = 4.52$, $\rho$ = 7.8, c/a = 1.5&
$a = 4.78, c = 2.948$, $\rho$ = 7.79, c/a = 0.617\\
\hline
\end{longtable}
\end{center}

\clearpage
\begin{figure}
\centering
\includegraphics[width=0.3\textwidth]{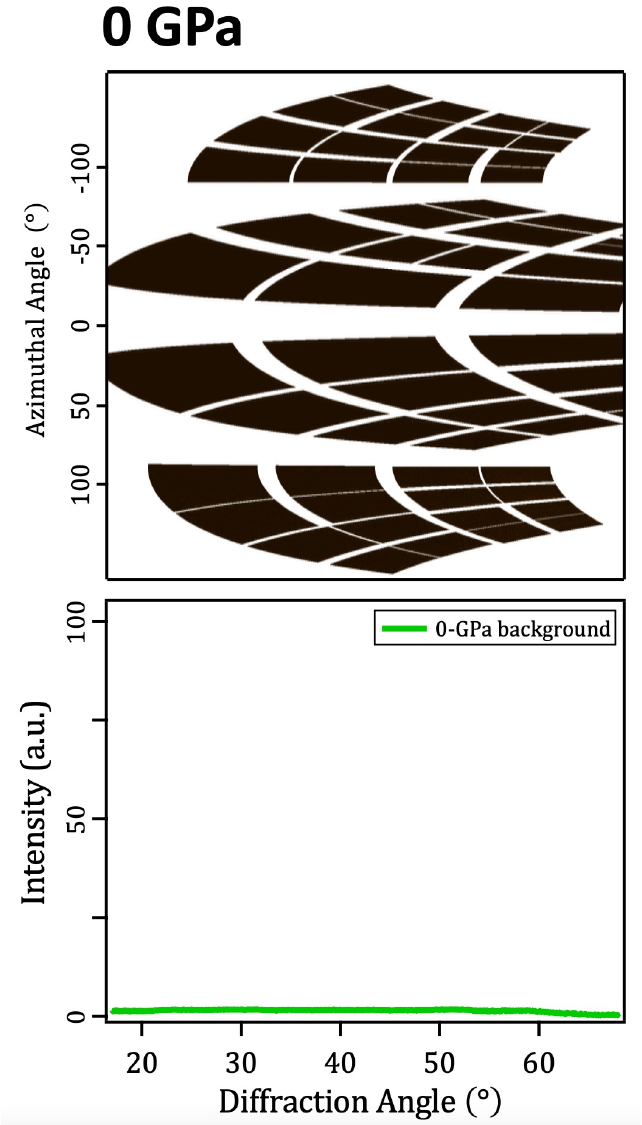}
\caption{\textbf{$\vert$ Establishing detector background.} The $2$-D (top) and azimuthally averaged (bottom) diffraction signal from an uncompressed Zr sample (0 GPa). The lack of any Bragg diffraction is a result of the highly-oriented nature of the starting sample. For this shot the XFEL X-rays were timed to probe the sample before the shock arrived at the Zr, therefore any small contribution from laser plasma X-rays is recorded on the detectors as a background level for the high pressure shots (e.g., Fig. \ref{fig:Fig_2}).}
\label{fig:SM_CSPAD}
\end{figure}

\section{\label{sec:poly}Polycrystalline Zr data}
\noindent

\begin{figure}[!ht]
\centering
\includegraphics[width=\textwidth]{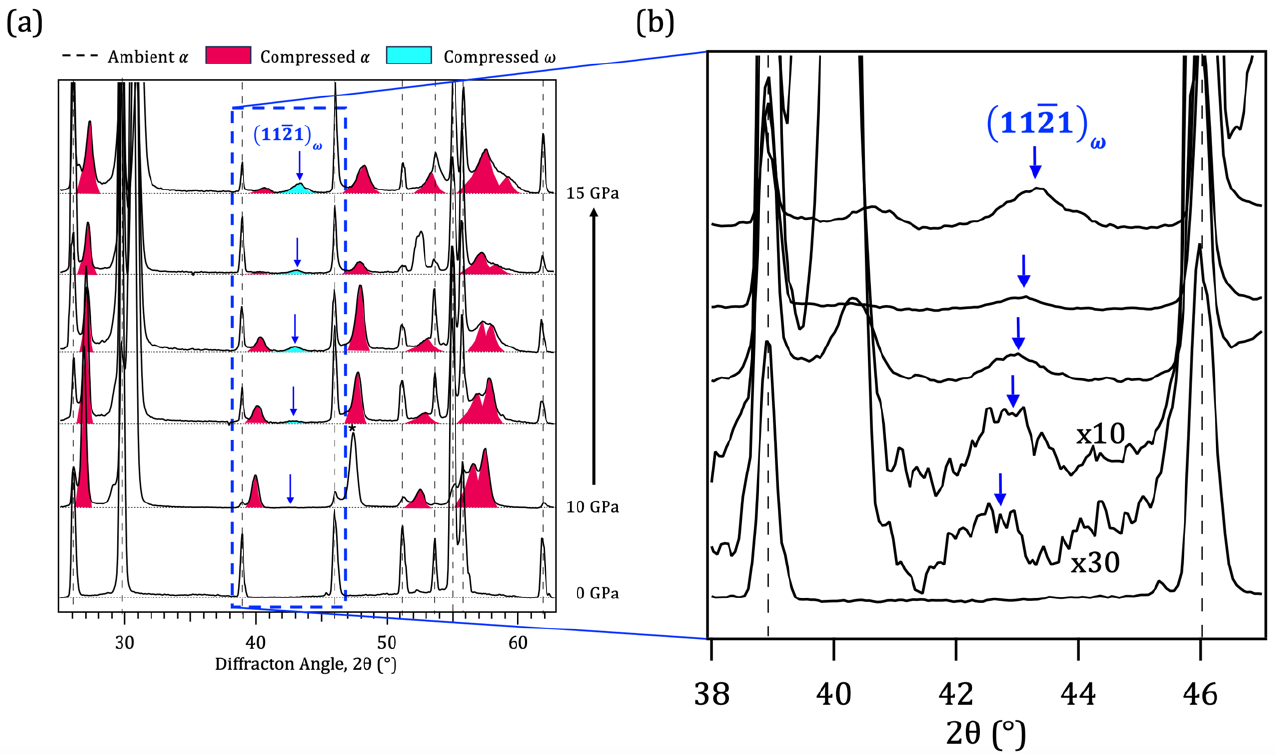}
\caption{\textbf{$\vert$ Diffraction data from shock-compressed polycrystalline Zr samples.} \textbf{a.} Waterfall plot of azimuthally-averaged diffraction intensities as a function of diffraction angle for poly-Zr shocked to $\sim$10 $\rightarrow$ 15 GPa [regime(ii)] -- comparable to the pressure range in Fig. \ref{fig:Fig_2}. All volume-integrated XRD data was taken during shock transit within the Zr layer, and therefore each trace contains a contribution from the ambient pressure $\alpha-$Zr volume ahead of the shock front (vertical dashed lines). Compressed $\alpha-$Zr peaks are labelled with an asterix. In contrast to the single crystal data in Fig. \ref{fig:Fig_2}, the polycrystalline samples exhibit no evidence of a disordered background (see dotted horizontal lines). \textbf{b.} Magnified view shows the emergence of the $\{11\bar{2}1\}_{\omega}$ peak.
\label{fig:Poly_Zr}
}
\end{figure}

Shock compression experiments along with {\it in-situ} x-ray diffraction were conducted for polycrystalline Zr samples in the same pressure ranges as discussed in the main text. A waterfall plot of the azimuthally averaged intensities for the ambient sample and shock compressed Zr in intermediate pressure regime [regime (ii)] of 10 -- 15 GPa is shown in Fig.~\ref{fig:Poly_Zr}(a). X-ray diffraction was recorded while the shock was transiting the Zr sample, which resulted in a diffraction contribution from both the compressed Zr, and the uncompressed volume ahead of the shock front. The location of the ambient pressure peaks have been marked with dashed vertical lines. The compressed $\alpha-$peaks have been labelled with an asterix ($^*$) and the $\omega-$phase has been labelled with a blue arrow. Compared to polycrystalline samples, {\it in-situ} x-ray diffraction during shock compression of single crystal Zr  exhibits a diffuse background superimposed on top of the crystalline diffraction, which reduces in magnitude as the shock pressure is increased (Fig.~\ref{fig:Fig_2}\textbf{a}).
While there have been no systematic experimental studies relating the shock response and phase transition pathways along various high symmetry directions of hcp metals, orientational anisotropy in the transition pathway and orientation relationships has been reported in large scale molecular dynamics simulations for titanium shocked along the $[0001], [10\bar{1}0])$ and $[11\bar{2}0]$ directions\cite{ZONG2014a}. We hypothesize a similar orientational dependence on the phase transition OR for Zr. This dependence could be related to the different response of the single crystal and polycrystalline response of Zr under shock compression. This question will be answered in future planned experiments.

\begin{figure}[!ht]
\centering
\includegraphics[width=\textwidth]{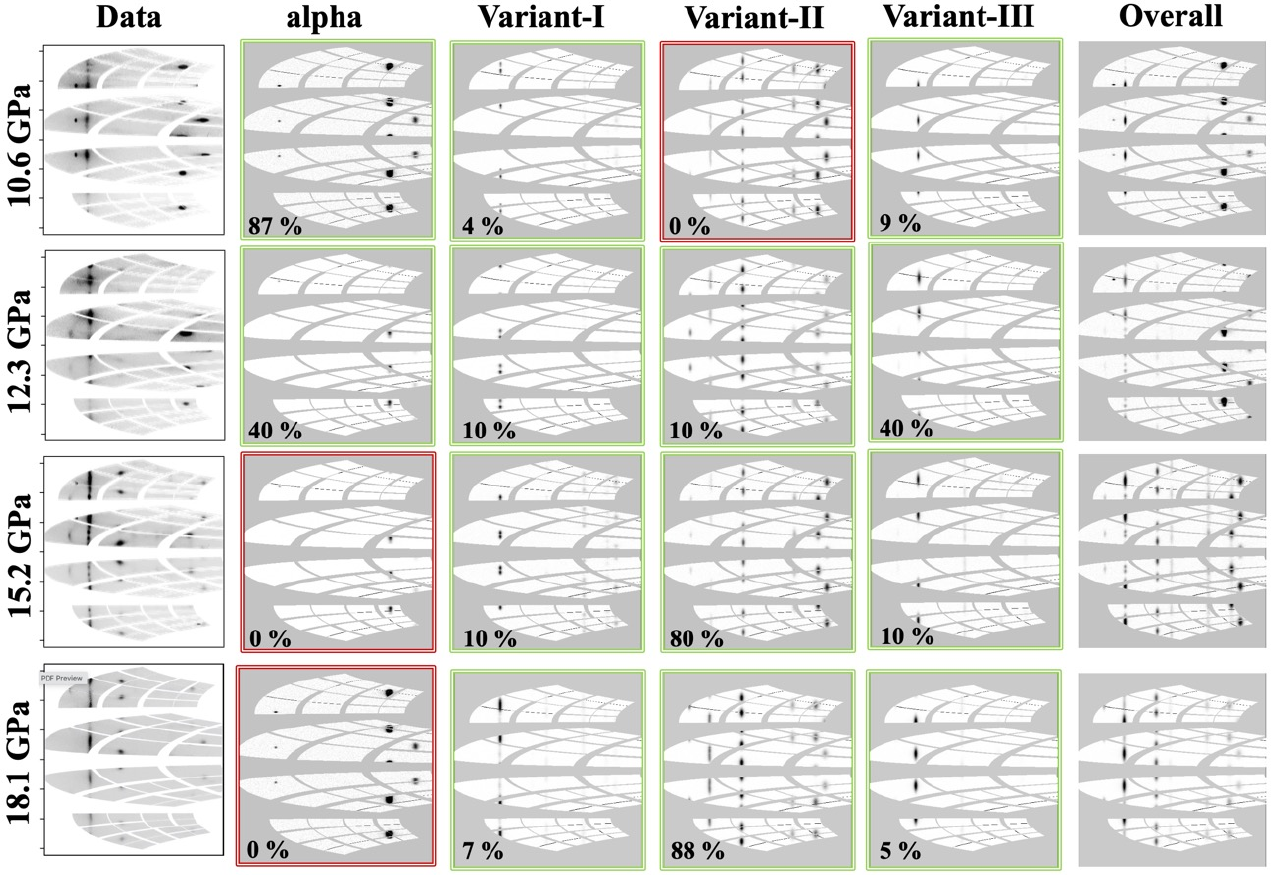}
\caption{Breakdown of the simulated diffraction pattern presented in Fig.~\ref{fig:Fig_2}\textbf{a}-\textbf{d} as a weighted sum of diffraction from the $\alpha-$phase, and variant I, variant II and variant III orientations of the $\omega-$phase. Left most column presents the same experimental data as Fig.~\ref{fig:Fig_2}\textbf{a}-\textbf{d} respectively. The \% weight indicated in the figure correspond to the phase fractions. Red box indicates no contribution from the particular phase or orientation variants.
\label{fig:phase_fraction}
}
\end{figure}

\begin{figure}[t!]
\centering
\includegraphics[width=1\textwidth]{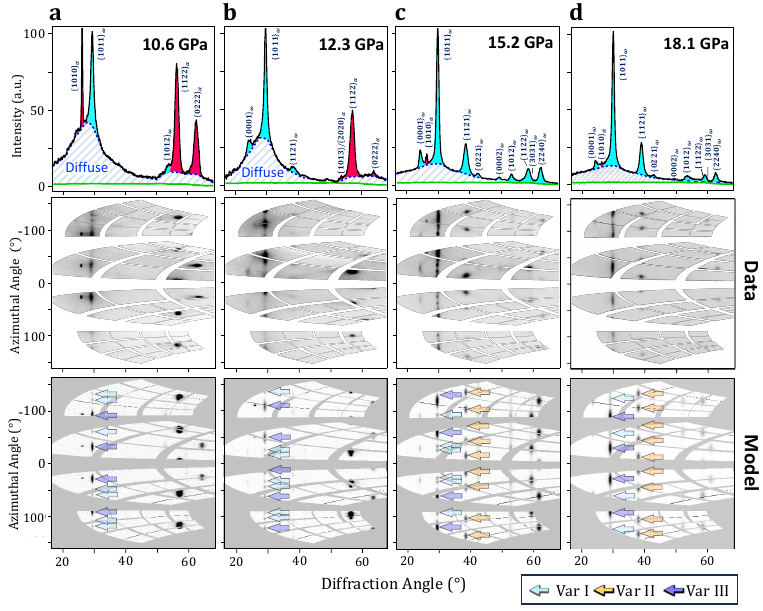}
\caption{\label{fig:Fig_2_old}
\textbf{$\vert$ Evolution of diffraction signal as a function of shock pressure.} The experimentally measured 1D azimuthally averaged (top) and full 2D diffraction signal (middle). The forward model predictions are shown in the bottom panel. \textbf{a.} At 10.6 GPa, the 2D and integrated diffraction signals show a combination of both $\alpha-$ (red filled) and $\omega-$phases (cyan filled). A large diffuse scattering background also accompanies the Bragg diffraction. \textbf{b.} At pressures above the onset of the phase transition [12.3 GPa], the $\{0001\}_{\omega}$ and $\{11\bar{2}1\}_{\omega}$ peaks start to emerge. A reduction in the diffuse background accompanies the emergence of these new peaks. \textbf{c.} An increase in shock pressure to 15.2 GPa leads to a further reduction in the phase fraction of the $\alpha-$phase and the diffuse scattering intensity. The new $\omega-$phase peaks observed in \textbf{b.}\ increase in intensity. \textbf{d.} At much higher pressures [18.1 GPa], the diffraction signal is composed almost entirely of the $\omega-$phase, with a small amount of diffraction from the compressed $\alpha-$phase, pointing to near complete conversion of the $\alpha \rightarrow \omega$ phase transition. There is a substantial reduction in the diffuse background signal and further evolution in the azimuthal intensity distribution.}
\end{figure}

\clearpage
\section{\label{sec:pair-correlation} Real-space structure of the partially disordered material in molecular dynamics simulations}
\begin{figure}[!ht]
\centering
\includegraphics[width=\textwidth]
{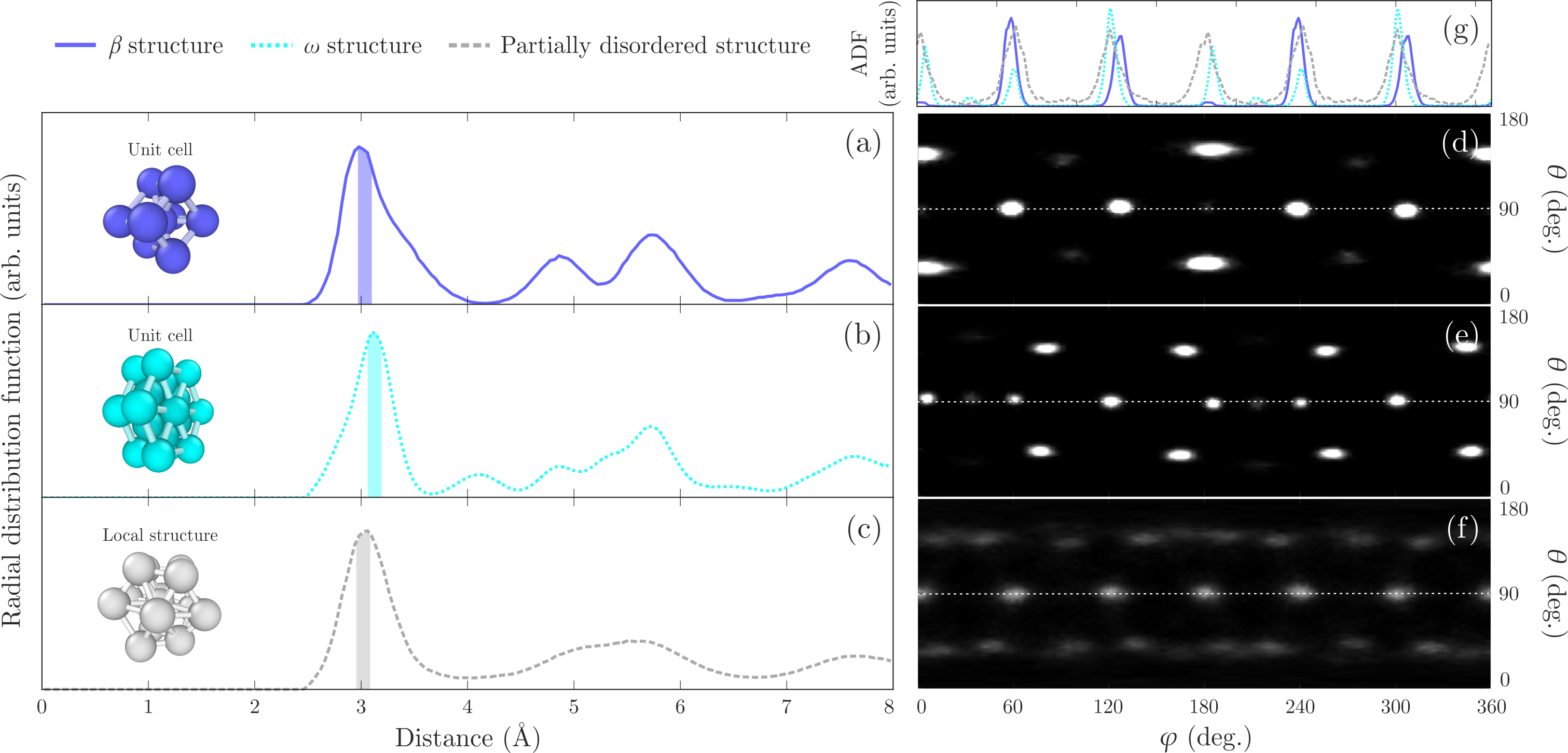}
\caption{\textbf{$\vert$ Pair-correlation functions for equilibrated $\beta$, $\omega$, and partially disordered material in Zr simulated under the Zong potential~\cite{ZONG2019} compressed to 14~GPa.} (a-c) Standard one-dimensional radial distribution functions (RDFs) calculated out to 8~\AA~separation. (d-f) Angular distribution functions (ADFs) for select $\beta$, $\omega$, and partially disordered nanoclusters ($N_{\text{atom}}\sim150)$, calculated at the separation indicated by the shaded bars in (a-c), time-averaged over at least 1~ps. Direction $\theta=0^\circ$ corresponds to the shock direction; $\theta=90^\circ$ is the `equatorial plane' containing directions normal to the shock. (g) Lineouts around the equatorial plane for the $\beta$, $\omega$, and partially disordered nanoclusters, with signal levels scaled arbitrarily for ease of comparison.
\label{fig:pair-correlation}
}
\end{figure}

The nanoclusters of noncrystalline material that form between the $\omega$ grains have an atomistic structure differing from that of the prevailing $\omega-$phase and residual $\beta-$phase. While the structure is not perfectly crystalline, it does exhibit a degree of translational and orientational order (hence our referring to this structure as `partially' disordered). Figs.~\ref{fig:pair-correlation}(a-c) show radial distribution functions (RDFs) for $\beta$, $\omega$, and partially disordered nanoclusters. The disordered structure exhibits translational order, though the correlation between its atomic separations is weaker than in the $\beta$ or $\omega$ crystalline phases. Figs.~\ref{fig:pair-correlation}(d-f) also shows the angular distribution functions (ADFs), defined as the probability of finding two atoms at a given fixed distance and separated by a displacement vector with direction $(\sin\theta\cos\varphi,\sin\theta\sin\varphi,\cos\theta)$ [with (0,0,1) being the shock direction]. We evaluate each structure's ADF at a separation corresponding to the first maximum in its respective RDF. We observe that the partially disordered structure shows strong orientational order despite it being noncrystalline. The strongly directional bonding within the partially disordered structure also manifests in its highly anisotropic structure factor, shown in Fig.~\ref{fig:structure_factors}.

\section{\label{sec:structure_factors} Reciprocal-space structure of the partially disordered material in molecular dynamics simulations} 
\begin{figure}[!ht]
\centering
\includegraphics[width=0.75\textwidth]
{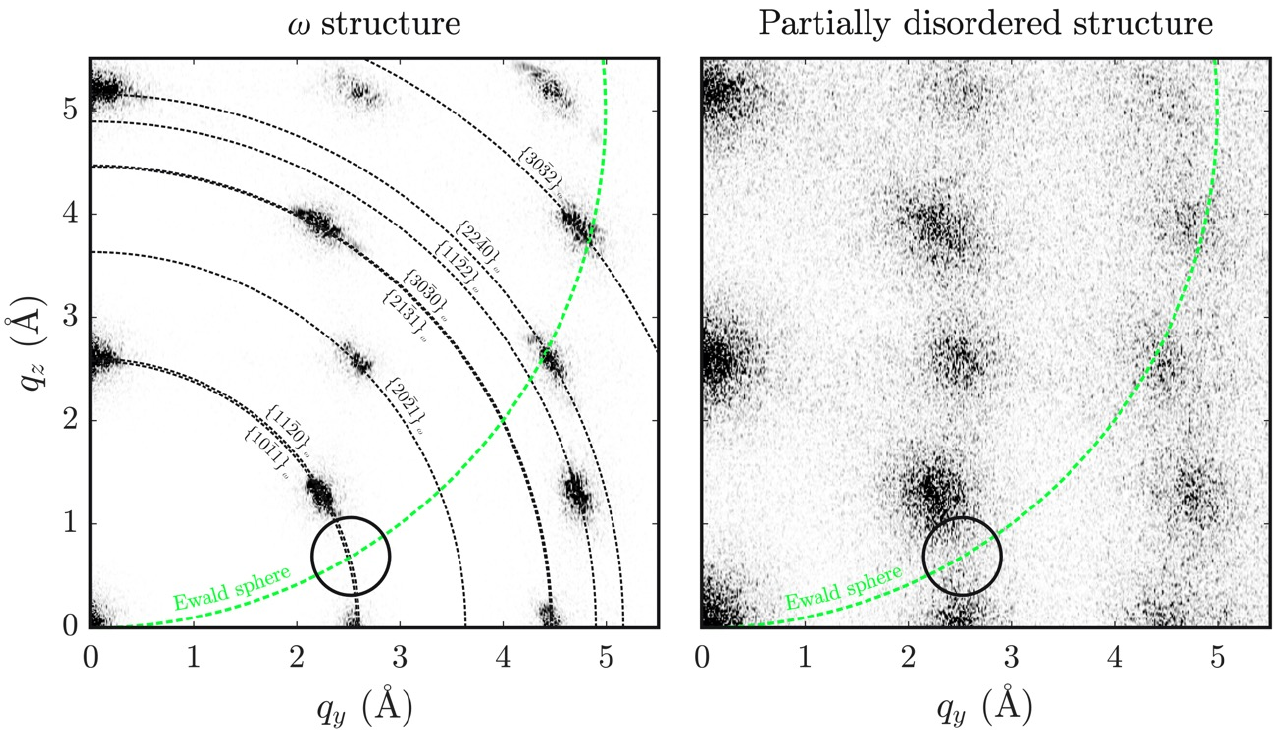}
\caption{\textbf{$\vert$ Comparison of the structure factors in a high-symmetry plane for the $\omega$ nanocrystal and the partially disordered nanoclusters (left and right, respectively) that form during shock-compression of Zr simulated under the Zong potential~\cite{ZONG2019}.} The cross-section of the viewing plane with the Ewald sphere (corresponding to a photon energy of 10~keV) and with the ideal Polanyi spheres for the $\omega$ structure are shown by dashed green and black lines, respectively. The black circle marks the region of reciprocal space from which the diffuse scattering signal (centered on $2\theta\sim30^\circ$) originates.
\label{fig:structure_factors}
}
\end{figure}

In Fig.~\ref{fig:structure_factors}, we show the squared modulus of the structure factor of the $\omega$ nanocrystal and the partially disordered nanoclusters, calculated in the high-symmetry $q_yq_z$ plane. The structure factor of the $\omega$ grains exhibit maxima at positions expected from their known variant II orientations. The structure factor of the partially disordered material also shows clear maxima at locations that coincide with those of the $\omega-$phase structure factor. We note that the greater width of the disordered material's maxima is partly attributable to it forming in relatively small nanoclusters. The similar forms of the two structure factors suggest that the $\omega$ nanocrystal and the partially disordered nanoclusters contribute to the diffraction pattern in qualitatively similar ways. In Fig.~\ref{fig:structure_factors}, we also indicate the region of reciprocal space from which the diffuse diffraction signal arises. This signal comes from the tenuous cloud of intensity surrounding the $\{10\bar{1}1\}_\omega$ and $\{11\bar{2}0\}_\omega$ reciprocal lattice vectors (and their equivalent maxima in the partially disordered material) that marginally miss the Bragg condition.

\clearpage
\section{\label{sec:beta}Intermediate bcc $\beta-$phase\label{sec:bcc}}
\begin{figure}
\centering
\includegraphics[width=0.8\textwidth]{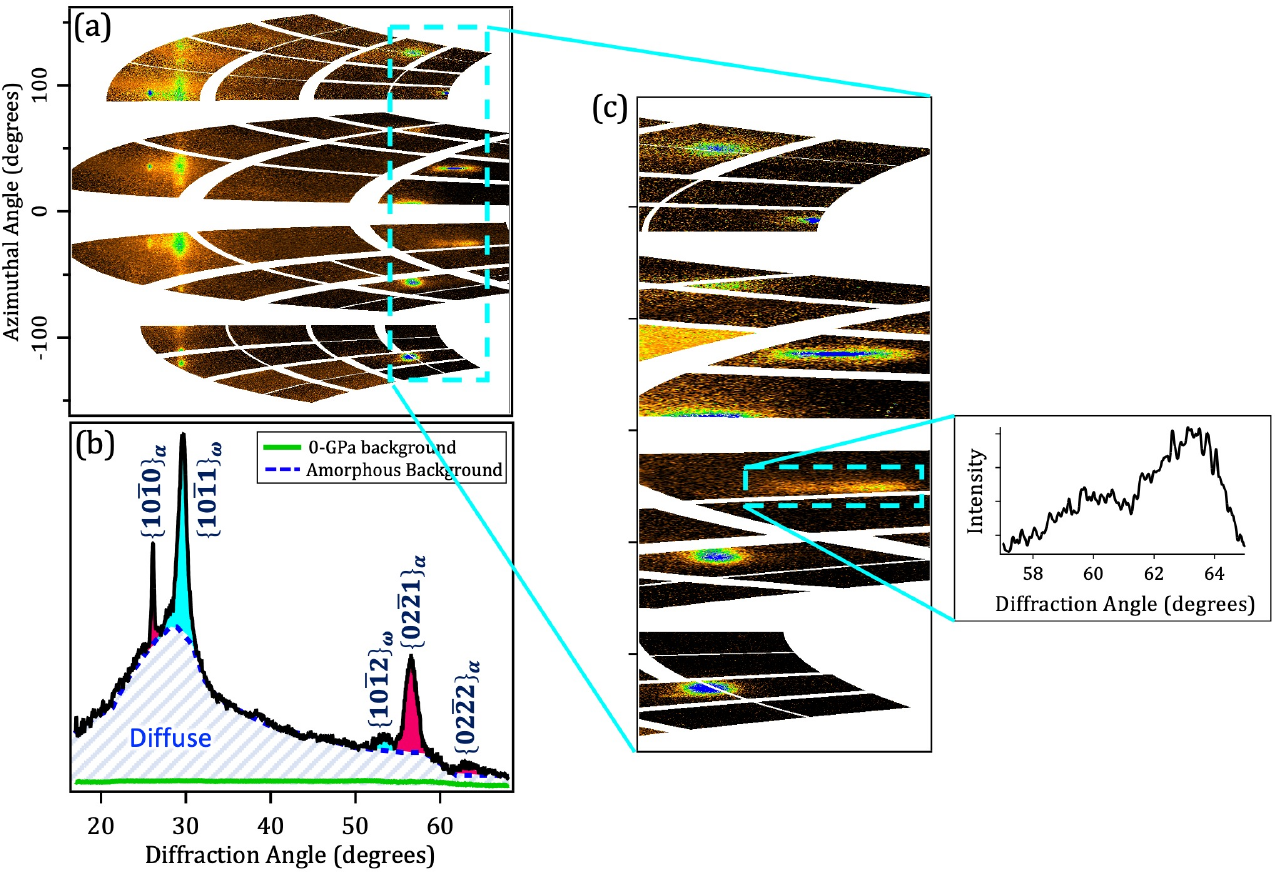}
\caption{\textbf{$\vert$ X-ray diffraction data taken at peak shock pressure of 10.4 GPa (shot 96).} \textbf{a.} 2D diffraction signal in the $2\theta-\phi$ space. \textbf{b.} Azimuthally averaged lineout showing the presence of sharp Bragg diffraction together with the diffuse scaatering signal. This was a repeat shot for the one shown in Fig.~\ref{fig:Fig_1} and Fig.~ \ref{fig:Fig_2_old}a. The demonstrates excellent reproducibility in the observations of texture and enhanced disordered background. \textbf{c.} Inset shows the high $2\theta$ region. A lineout for one of the $\{02\bar{2}2\}_{\alpha}$ peaks is also shown. This lineout shows two peaks separated in $2\theta$. The azimuthal location of the first peak at $\sim 60^{\circ}$ is consistent with the presence of an intermediate bcc $\beta-$phase. We discuss this further in section~\ref{sec:bcc}.\label{fig:s96}
}
\end{figure}
An intermediate bcc $\beta-$phase was initially hypothesized in Ref.~\cite{UZ1973} to explain the experimentally observed OR within recovered samples. Zong et al. \cite{ZONG2019,Zong2020} also reported a metastable $\beta-$phase in molecular dynamics simulations of shock-compressed single crystal Zr along the $c-$axis using interatomic potentials derived from a machine learning method. Finally, Armstrong et al. \cite{Armstrong2021} experimentally reported this phase in dynamic compression experiments on pure polycrystalline Zr using {\it in-situ} x-ray diffraction. 

Our MD simulations also show the transition of the hcp $\alpha-$phase to the bcc $\beta-$phase via rapid atomic shuffling in the $(0001)_\alpha$ plane [the Burgers mechanism~\cite{Burgers1934}] immediately behind the phase-transition front. After that, most of the transient $\beta$ phase decays into the stable $\omega$ phase via further in-plane shuffling (to produce variant II) or out-of-plane shuffling (variant I). However, the $\beta-$phase dominates for only a few picoseconds, and rapidly decays to a prohibitively small mass fraction $(\sim5\%)$.

\begin{figure}[!ht]
\centering
\includegraphics[width=0.9\textwidth]{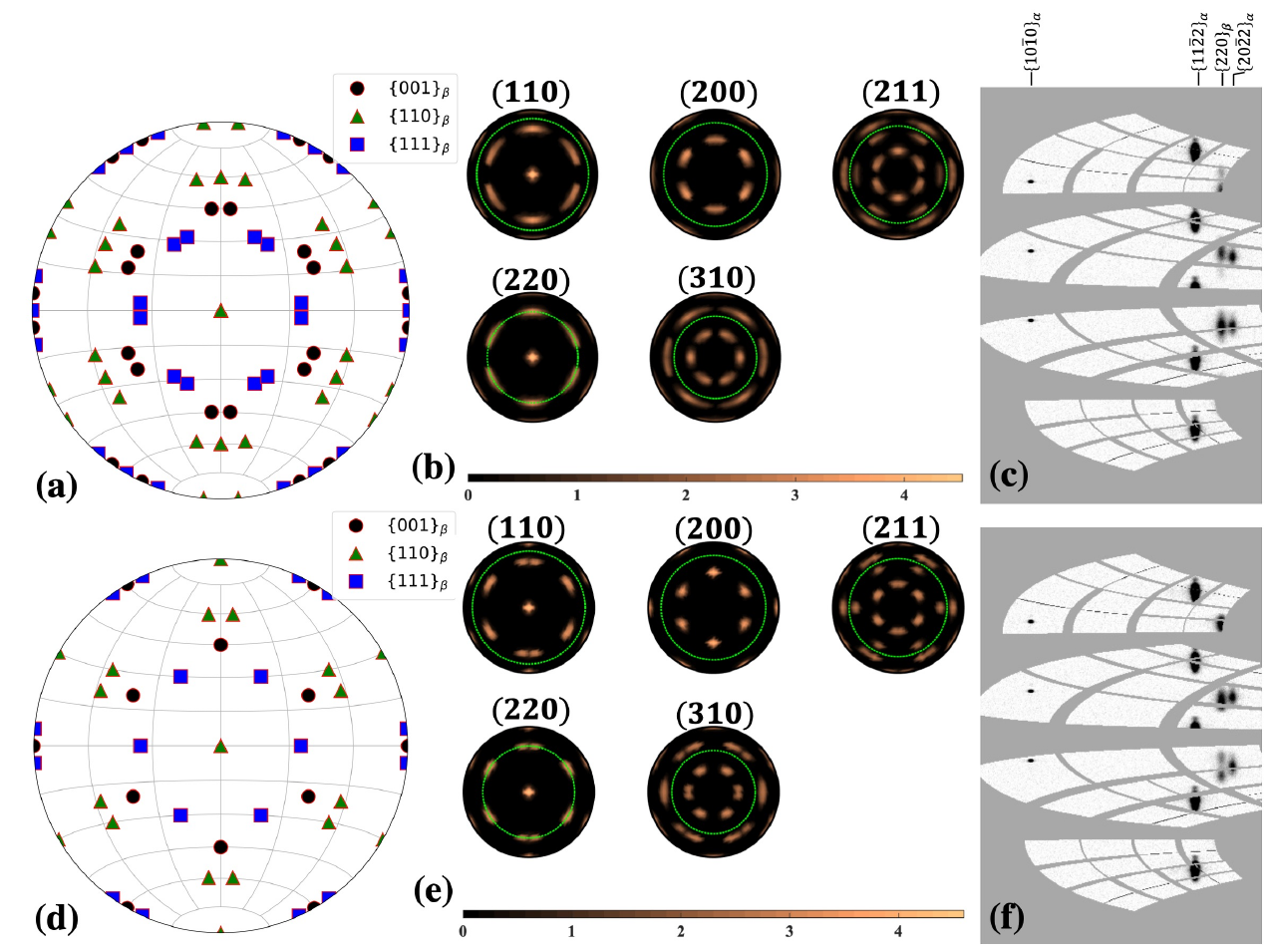}
\caption{Pole figures for \textbf{a.} a few important planes resulting due to the Burger's orientation relationship between the hexagonal closed pack $\alpha$ and bcc $\beta-$phase \textbf{b.} a unimodal orientation distribution of the $\beta-$phase resulting from the Burger's OR \textbf{c.} resulting diffraction signal due to the Burger's OR on the CSPAD detectors \textbf{d.} a few important planes resulting due to the Mao-Bassett-Takahashi (MBT) orientation relationship between the hcp
$\alpha-$ and bcc $\beta-$phase \textbf{e.} a unimodal orientation distribution of the $\beta-$phase resulting from the MBT OR and \textbf{f.} resulting diffraction signal due to the MBT OR on the CSPAD detectors. The green circles represent the orientations which meet the Bragg's conditions and observable in our experiments. If an intermediate $\beta-$phase was formed following the Burger's OR, then the $(220)$ planes should show a strong diffraction signal. Since this is not observed in our experiments, we conclude the absence of an intermediate bcc $\beta-$phase in the $\alpha\rightarrow\omega$ transition.
\label{fig:beta}
}
\end{figure}

We checked the consistency of our data with the presence of an intermediate bcc $\beta-$phase persisting in the \textit{ns} timescale. The orientation relationship between the hexagonal close-packed $\alpha$ and bcc $\beta-$phase in Zr has been experimentally determined to be the Burger's OR ($(0001)_{\alpha}~ \vert\vert~ (110)_{\beta}$ and $[2\bar{1}\bar{1}0]~ \vert \vert~ [\bar{1}11]_{\beta}$) \cite{Burgers1934} with a total of $6$ orientation variants. The stereographic projection of some important low index planes formed due to the Burger's OR for the starting orientation of $\alpha-$Zr in our experiment is shown in Fig.~\ref{fig:beta}a. To search for evidence of a bcc $\beta-$phase in our data we considered the peak labelled as $\{10\bar{1}1\}_{\omega}$ in Figs.~\ref{fig:Fig_2}\textbf{a}, \ref{fig:s96} as originating from the $\{110\}$ plane of the intermediate $\beta-$phase. This fixes the lattice parameter and density of the $\beta-$phase to a value of $7.31$ g~cm$^{-3}$. The calculated lattice parameter and the known ORs let us compute the diffraction signal originating from the $\beta-$phase.
We show the pole figure for unimodal orientation distribution centered around the $6$ orientations generated by the Burger's OR in Fig.~\ref{fig:beta}b. We represent the region of the pole figure observable in our experiments by the green overlays. Finally, we show the expected diffraction signal on the CSPAD detectors in Fig.~\ref{fig:beta}c. Our model predicts no diffraction from the $\{110\}_{\beta}$ planes but observable diffraction signal from the $(220)_{\beta}$ reflection. We predict the $(220)_{\beta}$ diffraction signal at the same azimuthal angle as the $(02\bar{2}2)_{\alpha}$ reflection. In addition, the expected $2\theta$ location for the two reflections is close. We look closer at this region of the CSPAD detectors in Fig.~\ref{fig:s96}inset. The $2\theta$ lineout for one of the $(02\bar{2}2)_{\alpha}$ shows evidence for two peaks. This observation is consistent with a tiny volume fraction of the intermediate $\beta-$phase. We note that this feature is repeatable (see Fig.~\ref{fig:Fig_1}\textbf{b}). While the above observation suggests that we are measuring diffraction from an intermediate $\beta-$phase as seen in our MD simulations, we do not believe that the signal level or detector coverage is high enough to be conclusive. We plan to repeat these measurements with better detector coverage in future experiments.

We also checked the consistency of other hcp-bcc ORs with our data. We list the different ORs we considered when interpreting our data in Table~\ref{Table:OR_bcc}. For example, we show the pole figures for unimodal orientation distribution centered around the $3$ orientations generated by the MBT OR in Fig.~\ref{fig:beta}e, with the green overlays showing the region of the pole figure observable in our experiments. We present the expected diffraction signal on the CSPAD detectors in Fig.~\ref{fig:beta}f. The diffraction produced by this OR is very similar to Burger's OR. Therefore, we are unable to distinguish between the two.

We also note that while the Rong-Dunlop \cite{rong1984} and the Song Du-Sun \cite{Song2002} ORs, for hcp metal-carbon precipitates in bcc iron, produce an azimuthal intensity distribution matching our experiments for the currently labeled $(10\bar{1}1)_{\omega}$ reflection, and indistinguishable from our variant III, these ORs have never been experimentally observed or theoretically predicted in Zr. Since these ORs have only been observed in high strength steel for hcp carbide precipitates within the bcc ferrite matrix, we deem it highly unlikely that these ORs are present in the $\beta-$phase during the $\alpha\rightarrow\omega$ phase transition in Zr.

\section{\label{sec:var} Orientation Variants}

\begin{figure}[!ht]
\centering
\includegraphics[width=0.75\textwidth]{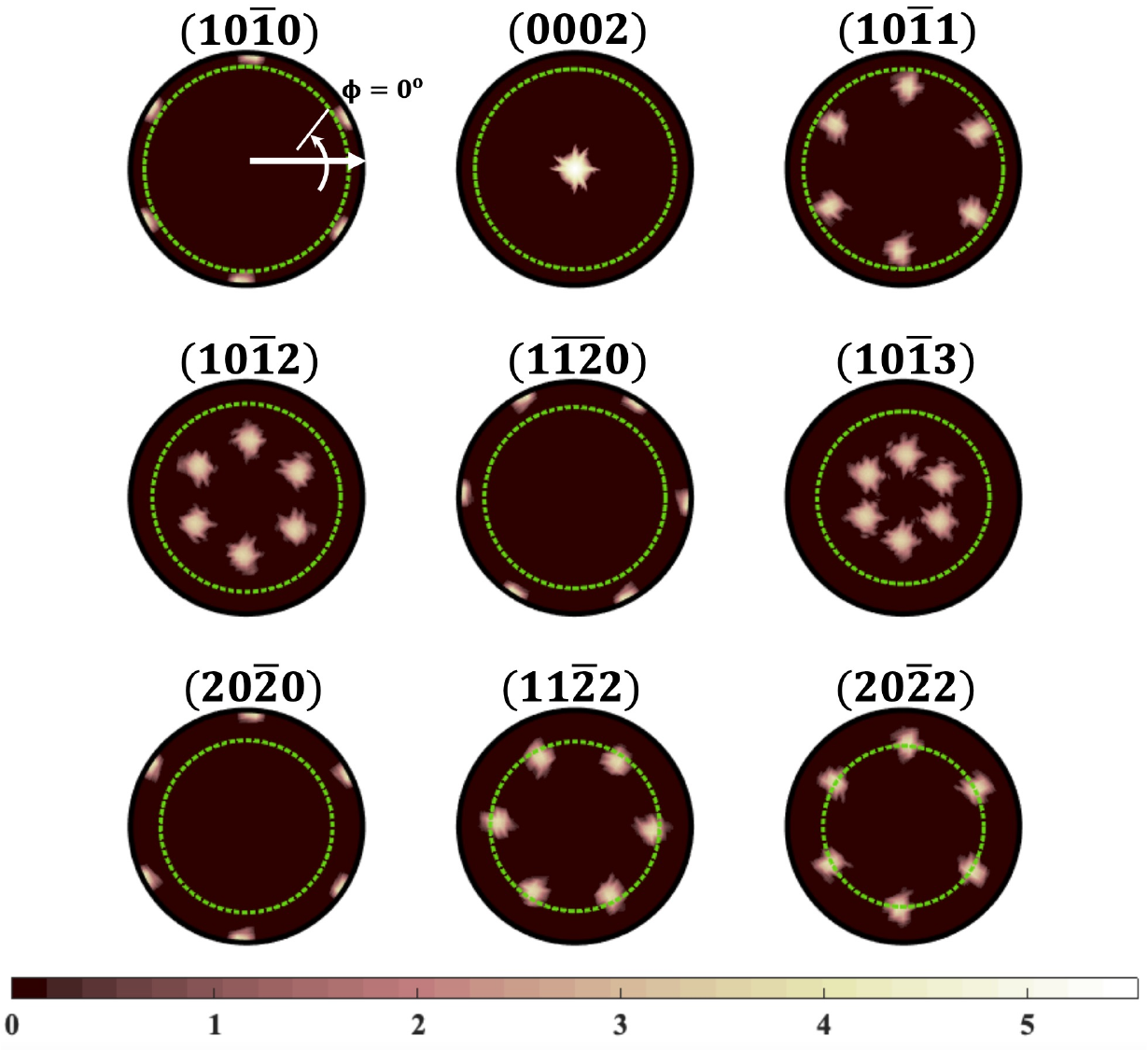}
\caption{Pole figures (log scale) for a unimodal orientation distribution of the $\alpha-$phase centered around the orientation of the ambient crystal, i.e. $c-$axis pointing in the direction of the x-rays and $a-$axis normal to it, pointing east. The observable rings in our experimental geometry for each reflection is shown by the dotted green circle. When the circle overlaps with non-zero intensity in the pole figure, then diffraction is observed for that reflection. The variation of intensity around the circle is reflected in the observed diffraction signal.
\label{fig:FM_alpha}
}
\end{figure}

\begin{figure}[!ht]
\centering
\includegraphics[width=0.75\textwidth]{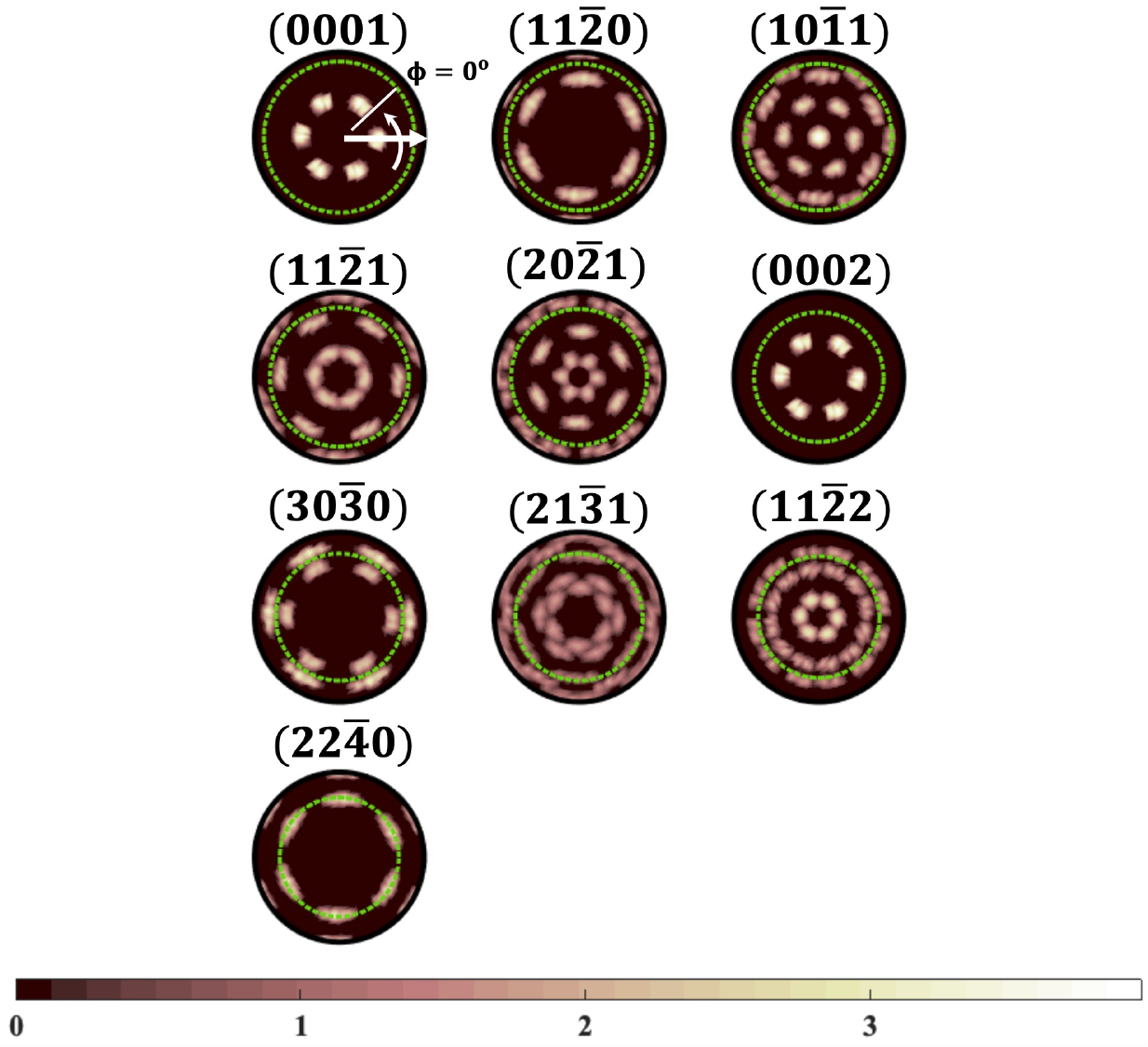}
\caption{Pole figures (log scale) for a unimodal orientation distribution of the $\omega-$phase centered around the orientations from the variant I (TAO-I) OR. The observable rings in our experimental geometry for each reflection is shown by the dotted green circle. When the circle overlaps with non-zero intensity in the pole figure, then diffraction is observed for that reflection. The variation of intensity around the circle is reflected in the observed diffraction signal.
\label{fig:FM_omega_var1}
}
\end{figure}

\begin{figure}[!ht]
\centering
\includegraphics[width=0.75\textwidth]{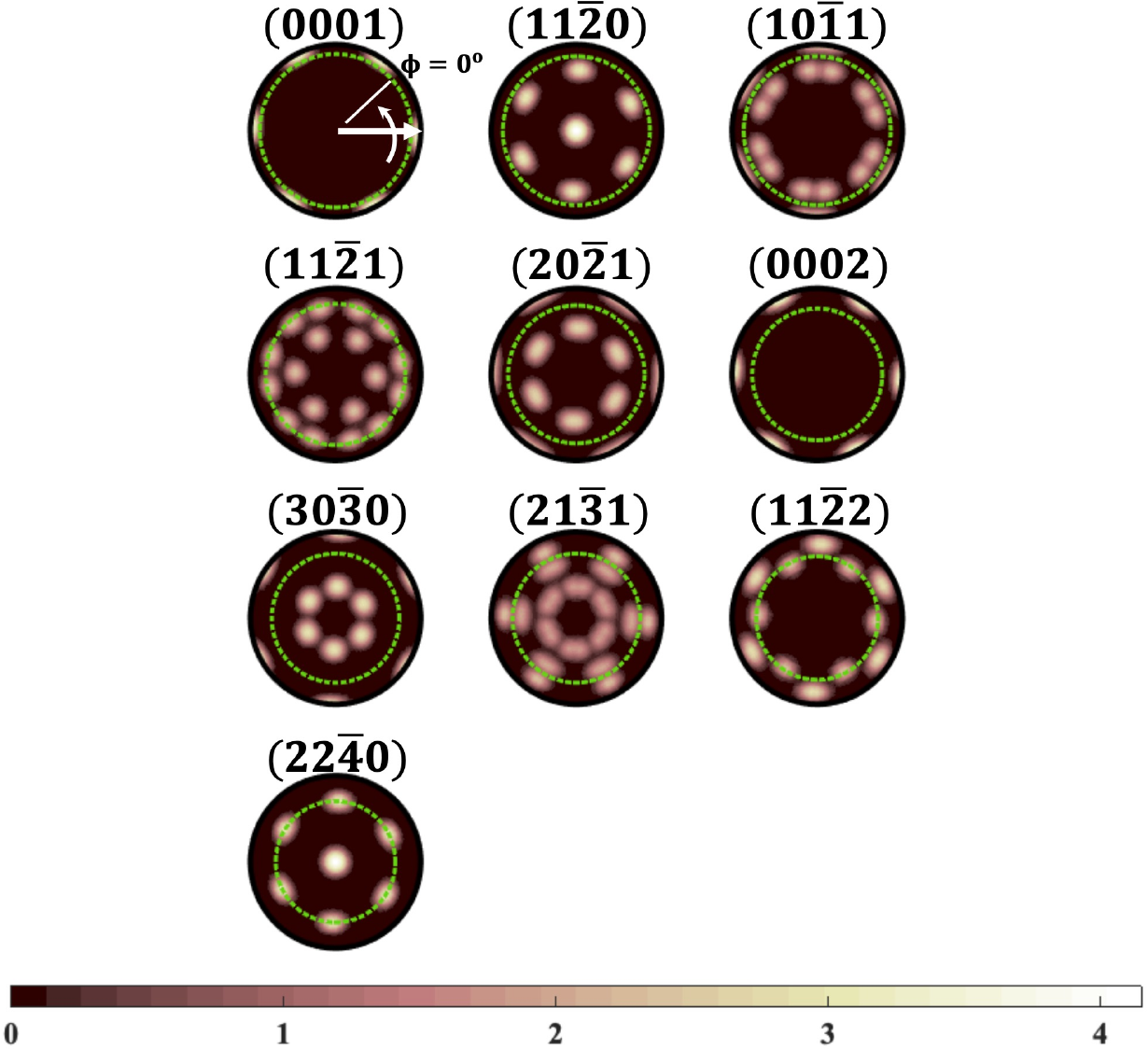}
\caption{Pole figures (log scale) for a unimodal orientation distribution of the $\omega-$phase centered around the orientations from variant II (Silcock) OR. The observable rings in our experimental geometry for each reflection is shown by the dotted green circle. When the circle overlaps with non-zero intensity in the pole figure, then diffraction is observed for that reflection. The variation of intensity around the circle is reflected in the observed diffraction signal.
\label{fig:FM_omega_var2}
}
\end{figure}

\begin{figure}[!ht]
\centering
\includegraphics[width=0.75\textwidth]{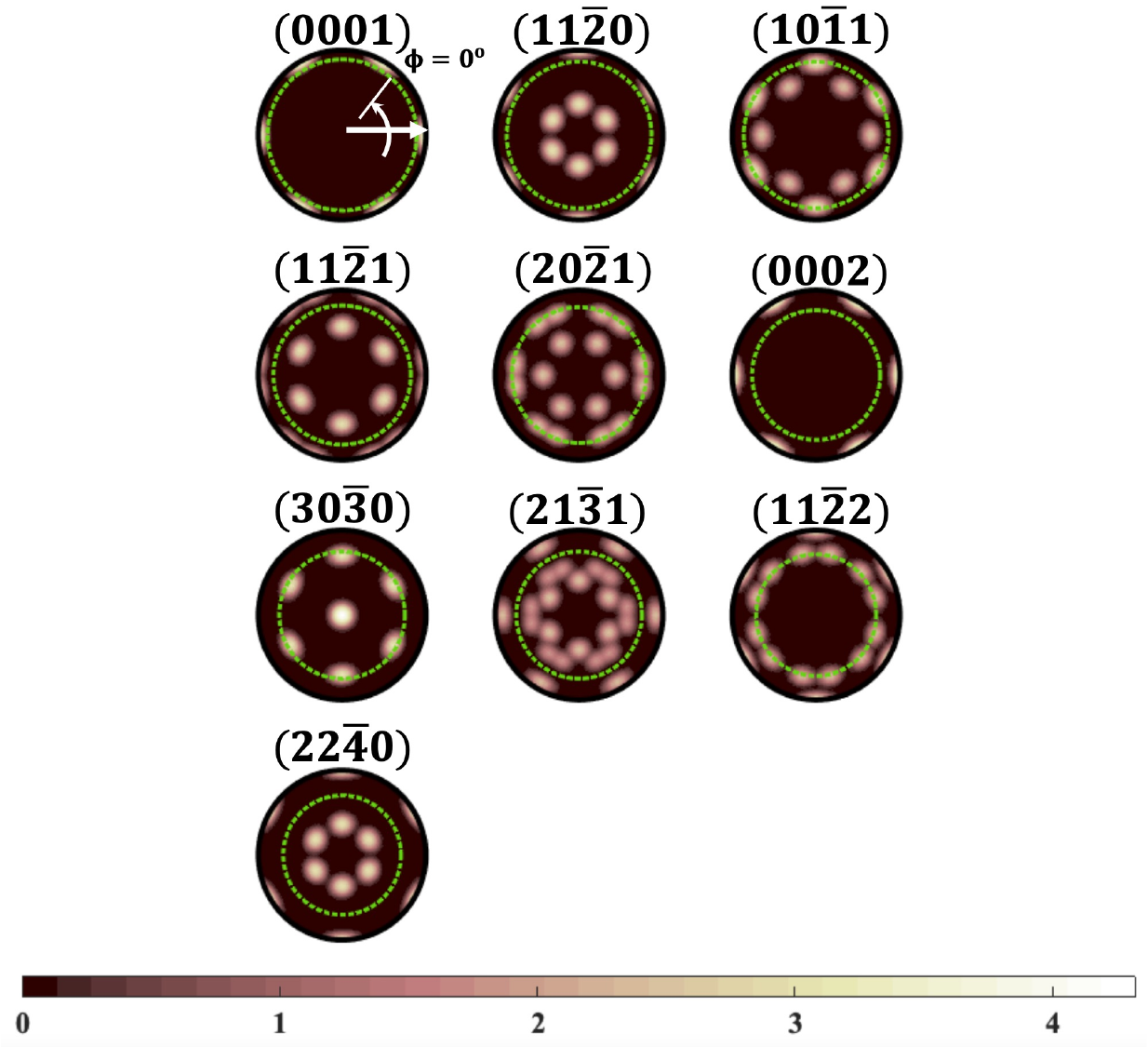}
\caption{Pole figures (log scale) for a unimodal orientation distribution of the $\omega-$phase centered around the orientations from variant III OR. The observable rings in our experimental geometry for each reflection is shown by the dotted green circle. When the circle overlaps with non-zero intensity in the pole figure, then diffraction is observed for that reflection. The variation of intensity around the circle is reflected in the observed diffraction signal.
\label{fig:FM_omega_var3}
}
\end{figure}

Each of the orientation relationships listed in Table~\ref{Table:OR} describes one particular change of basis between the parent $\alpha$ and the daughter $\omega-$phase. However, due to crystallographic symmetry, it is possible to form other $\omega-$phase with a different orientation. For example, the variant I OR described by $(0001)_{\alpha} \vert\vert (01\bar{1}1)_{\omega}$ and $[11\bar{2}0]_{\alpha} \vert\vert [10\bar{1}1]_{\omega}$ can also form such that $(0001)_{\alpha} \vert\vert (1\bar{1}01)_{\omega}$ and $[11\bar{2}0]_{\alpha} \vert\vert [\bar{1}011]_{\omega}$ etc. Therefore, when computing the diffraction signal resulting from the resultant $\omega-$phase, it is important to consider all the orientation variants due to a particular OR. In this section, we list the number of variants for each of the OR described above and show distribution of different crystallographically important planes and directions for each of the ORs. Given the transformation matrix, $\mathbf{T_{\alpha\rightarrow\omega}}$ between the parent and daughter basis, and the point group symmetry, $\mathbf{G}^{\alpha}$ and $\mathbf{G}^{\omega}$ of the two phases, the common symmetries between the two phases, $\mathbf{H}^{\alpha,\omega}$ is given by
\begin{equation*}
\mathbf{H}^{\alpha,\omega} = \mathbf{G}^{\alpha} ~ \cap ~ \mathbf{T_{\alpha\rightarrow\omega}}\mathbf{G}^{\omega}\mathbf{T_{\alpha\rightarrow\omega}}^{-1}.
\end{equation*}
The number of orientation variants, $N_{\alpha\rightarrow\omega}$ is then given by the following expression
\begin{equation*}
N_{\alpha\rightarrow\omega} = \frac{\vert \mathbf{G}^{\alpha} \vert}{\vert \mathbf{H}^{\alpha,\omega} \vert}.
\end{equation*}
Here, 
$\vert \mathbf{A}\vert$ represents the number of elements in group $\mathbf{A}$. Each orientation variant, $\alpha_{i}$ can be calculated from symmetry elements, $g_{i}^{\alpha}$ the parent phase point group, $\mathbf{G}^{\alpha}$. The orientation variants for each OR can be computed using the symmetry group of the parent phase. If $\mathbf{T_{\alpha\rightarrow\omega}}$ denotes the transformation matrix between the $\alpha$ and the $\omega-$phase given by the OR, then the orientation variants are the unique elements in the set, $\mathbf{T^{\prime}_{\alpha\rightarrow\omega}}$, given by
\begin{equation*}
\mathbf{T^{\prime}_{\alpha\rightarrow\omega}} = \mathbf{G}^{\alpha} \mathbf{T_{\alpha\rightarrow\omega}}.
\end{equation*}

\begin{figure}
\centering
\includegraphics[width=1\textwidth]{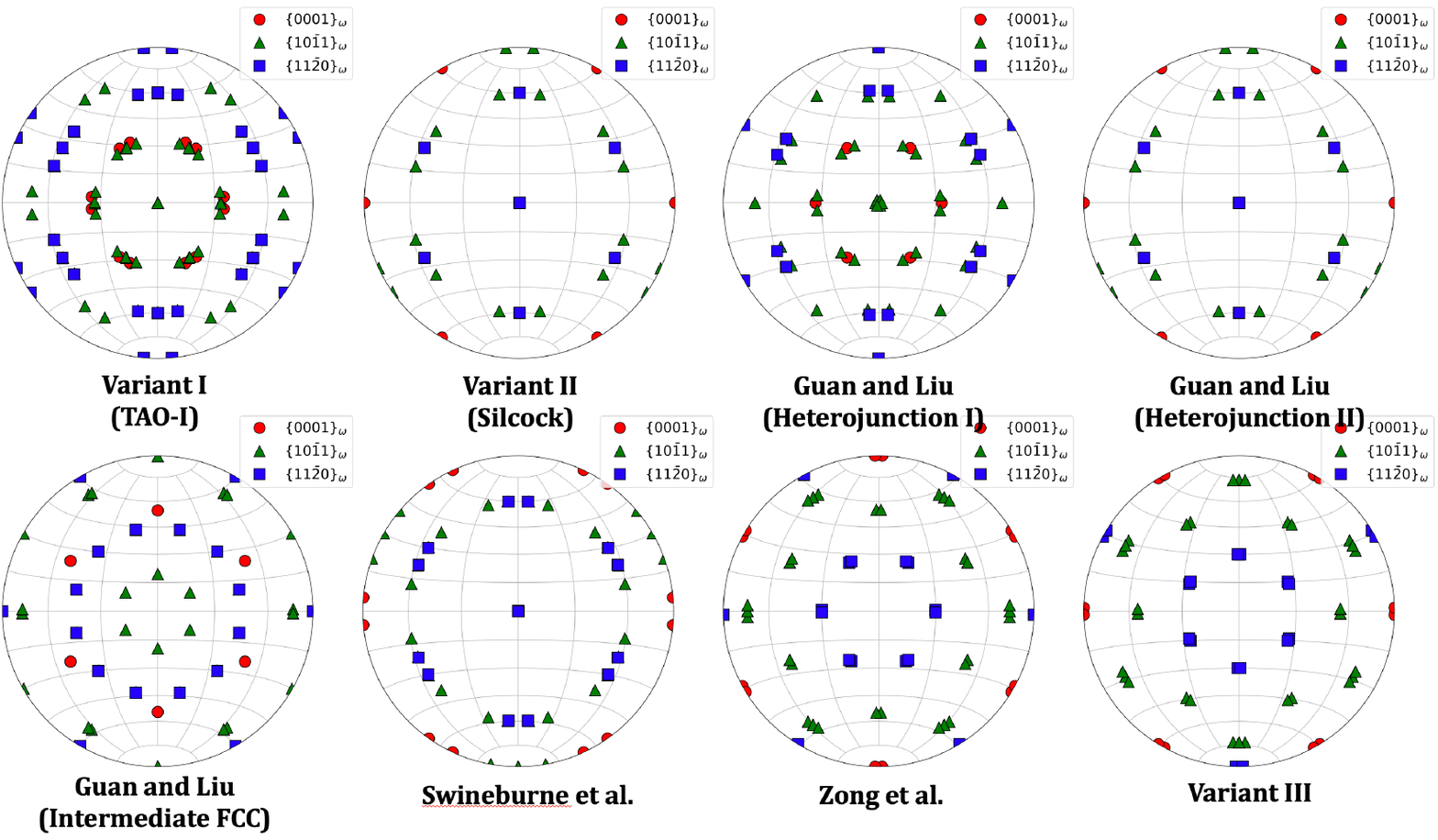}
\caption{Distribution of the $\{0001\}, \{10\bar{1}1\}$ and $\{11\bar{2}0\}$ planes of the $\omega-$phase for different ORs reported in experimental and theoretical studies. References for each of these studies can be found in Table~\ref{Table:OR}. The heterojunction I OR reported in Guan and Liu \cite{Guan2016} is reported to be equivalent to variant I OR. While the ORs are similar, they are not equivalent. Heterojunction II OR is identical to the variant II or Silcock OR. Similar distribution for the other crystallographic planes can be computed and used to calculated the expected diffraction signal on the CSPAD detectors. Direct comparison with observed diffraction can be used to identify the resultant orientation relationship between $\alpha$ and $\omega-$phases. \label{fig:SM_PFs}}
\end{figure}

The reader is referred to \cite{Cayron2007} for a more comprehensive description of the mathematical structure of orientation variants. All orientation variants of the different ORs in the literature and their respective pole figures were computed to compare with our experimental measurements. Fig.~\ref{fig:SM_PFs} presents the stereographic projections of a few crystallographically important planes of the $\omega-$phase for the different ORs.
\clearpage
\small{
\begin{longtable}{|p{0.3\textwidth}|p{0.2\textwidth}|}
	\caption{\textbf{$\vert$ Number of orientation variants for different $\alpha\rightarrow\omega$ orientation relationships}\label{Table:numvariants1}} \\
	\hline
	\textbf{Orientation relationship} & \textbf{\# variants} \\
	\hline
	$(0001)_{\alpha}~\vert\vert~(01\bar{1}1)_{\omega}$\newline
	$[11\bar{2}0]_{\alpha}~\vert\vert~[10\bar{1}1]_{\omega}$\newline
	(\textbf{variant I}) & 12 \\
	\hline
	$(0001)_{\alpha}~\vert\vert~(11\bar{2}0)_{\omega}$\newline 
	$[11\bar{2}0]_{\alpha}~\vert\vert~[0001]_{\omega}$\newline 
	(\textbf{variant II}) & 3 \\
	\hline
	$(0001)_{\alpha}~\vert\vert~(1\bar{1}00)_{\omega}$ \newline
	$[11\bar{2}0]_{\alpha}~\vert\vert~[11\bar{2}3]_{\omega}$\newline
	(\textbf{variant III}) & 6 \\
	\hline
	$(0001)_{\alpha}~\vert\vert~(1\bar{1}00)_{\omega}$\newline
	$[10\bar{1}0]_{\alpha}~\vert\vert~[11\bar{2}3]_{\omega}$ \newline
	closely related to \textbf{variant III} \cite{ZONG2014a}) & 6 \\
	\hline
	$(10\bar{1}0)_{\alpha}~\vert\vert~ (10\bar{1}1)_{\omega}$\newline
	$[0001]_{\alpha}~ \vert\vert~ [1\bar{2}10]_{\omega}$\newline
	(Swinburne et al. \cite{Swinburne2016}) & 6 \\
	\hline
	$(11\bar{2}2)_{\alpha}~\vert\vert~(1\bar{1}00)_{\omega}$\newline
	$[1\bar{1}00]_{\alpha}~\vert\vert~[11\bar{2}0]_{\omega}$\newline
	(closely related to \textbf{variant I (TAO-I)} \cite{Guan2016}) & 6 \\
	\hline
	$(10\bar{1}0)_{\alpha}~\vert\vert~(1\bar{1}00)_{\omega}$ \newline $[0001]_{\alpha}~\vert\vert~[11\bar{2}0]_{\omega}$\newline
	(\textbf{variant II (Silcock/Heterojunction II)} \cite{Guan2016}) & 3 \\
	\hline
	$(0001)_{\alpha}~\vert\vert~(02\bar{2}1)_{\omega}$\newline
	$[11\bar{2}0]_{\alpha}~\vert\vert~[2\bar{1}\bar{1}0]_{\omega}$\newline
	(\textbf{Guan and Liu (intermediate FCC)} \cite{Guan2016}) & 6 \\
	\hline
\end{longtable}
}
\begin{longtable}{|p{0.3\textwidth}|p{0.2\textwidth}| p{0.27\textwidth}|}
\caption{\textbf{$\vert$ Number of orientation variants for different $\alpha\rightarrow\beta$ orientation relationships}\label{Table:numvariants2}} \\
\hline
\centering
\textbf{Orientation relationship} & \textbf{\# variants} \\
\hline
$(110)_{\text{\small bcc}}~\vert\vert~(0001)_{\text{\small hcp}}$\newline
$[1\bar{1}1]_{\text{\small bcc}}~\vert\vert~[11\bar{2}0]_{\text{\small hcp}}$\newline
(\textbf{Burger's} \cite{Burgers1934}) & 6 \\
\hline
$(110)_{\text{\small bcc}}~\vert\vert~(0001)_{\text{\small hcp}}$\newline
$[1\bar{1}0]_{\text{\small bcc}}~\vert\vert~[10\bar{1}0]_{\text{\small hcp}}$\newline
(\textbf{Pitsch-Schrader} \cite{Pitsch1958}) & 3 \\
\hline
$(110)_{\text{\small bcc}}~\vert\vert~(0001)_{\text{\small hcp}}$\newline
$[001]_{\text{\small bcc}}~\vert\vert~[2\bar{1}\bar{1}0]_{\text{\small hcp}}$\newline
(\textbf{Mao-Bassett-Takahashi} \cite{Mao1967}) & 3 \\
\hline
$(110)_{\text{\small bcc}}~\vert\vert~(0001)_{\text{\small hcp}}$\newline
$[1\bar{1}2]_{\text{\small bcc}}~\vert\vert~[1\bar{2}10]_{\text{\small hcp}}$\newline
(\textbf{Gjönnes-Östmer} \cite{Gjonnes1970}) & 12 \\
\hline
$(110)_{\text{\small bcc}}~\vert\vert~(1\bar{1}01)_{\text{\small hcp}}$\newline
$[1\bar{1}1]_{\text{\small bcc}}~\vert\vert~[11\bar{2}0]_{\text{\small hcp}}$\newline
(\textbf{Potter} \cite{potter1973}) & 12 \\
\hline
$(111)_{\text{\small bcc}}~\vert\vert~(0001)_{\text{\small hcp}}$\newline
$[11\bar{2}]_{\text{\small bcc}}~\vert\vert~[2\bar{1}\bar{1}0]_{\text{\small hcp}}$\newline
(\textbf{Crawley-Milliken} \cite{Crawley1974}) & 2 \\
\hline
$(021)_{\text{\small bcc}}~\vert\vert~(0001)_{\text{\small hcp}}$\newline
$[100]_{\text{\small bcc}}~\vert\vert~[11\bar{2}0]_{\text{\small hcp}}$\newline
(\textbf{Rong-Dunlop} \cite{rong1984}) & 6 \\
\hline
$(010)_{\text{\small bcc}}~\vert\vert~(01\bar{1}1)_{\text{\small hcp}}$\newline
$[100]_{\text{\small bcc}}~\vert\vert~[2\bar{1}\bar{1}0]_{\text{\small hcp}}$\newline
(\textbf{Song-Du-Sun} \cite{Song2002}) & 6 \\
\hline
\end{longtable}

\begin{figure}[!ht]
\centering
\includegraphics[width=0.75\textwidth]{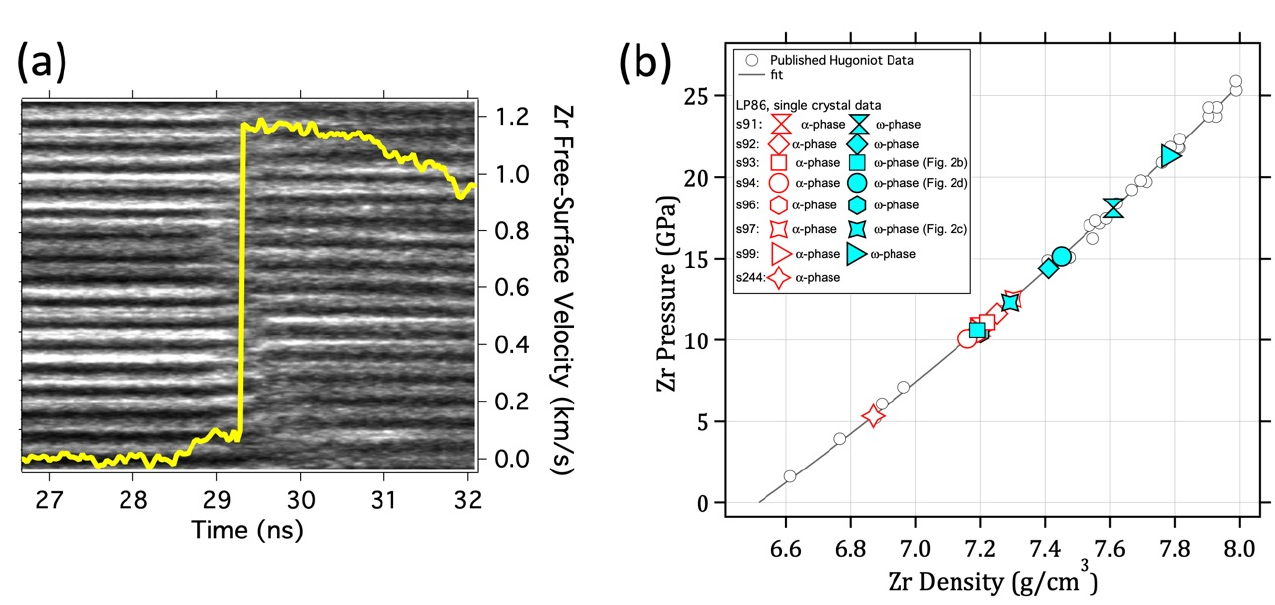}
\caption{\textbf{$\vert$ Velocimetry measurements and pressure determination. a.} A representative line-visar image taken over a 300 $\mu$m field of view \cite{Celliers2004}. The determined Zr-free surface velocity is shown as the yellow trace. This data was used to verify: (i) the relative timing of the x-ray probe and the compression wave, (ii) the temporal steadiness of the laser shock. \textbf{b.} Pressure-density Hugoniot data established from previous work (open black circles) \cite{al1981,walsh1957,mcqueen1970,rigg2014}. Density determined from x-ray diffraction data ($\alpha-$phase: red open symbols; $\omega-$phase: blue filled symbols). In total eight single crystal shots where taken. The diffraction patterns from shots 93, 94 and 97 are shown in Fig. \ref{fig:Fig_2}. The diffraction pattern for shot 96 is shown in Fig. \ref{fig:s96}. To determine pressure in our experiments, we fix the measured density of the $\alpha$ and $\omega-$phases to the fit through previously determined Hugoniot data (grey curve). Below shock pressures of 10.2 GPa only compressed $\alpha-$peaks are observed in the X-ray diffraction data. The highest pressure we observe compressed $\alpha-$Zr is 12.3 GPa. For shock pressures below this value, the $\omega-$Zr pressure and density within the mixed phase assembly is higher than the $\alpha-$phase density.
	\label{fig:VISAR}
}
\end{figure}

\section{\label{sec:TAO}TAO-I mechanism (variant I) during $\alpha\rightarrow\omega$ transformation}
\begin{figure}[!ht]
\centering
\includegraphics[width=0.8\textwidth]{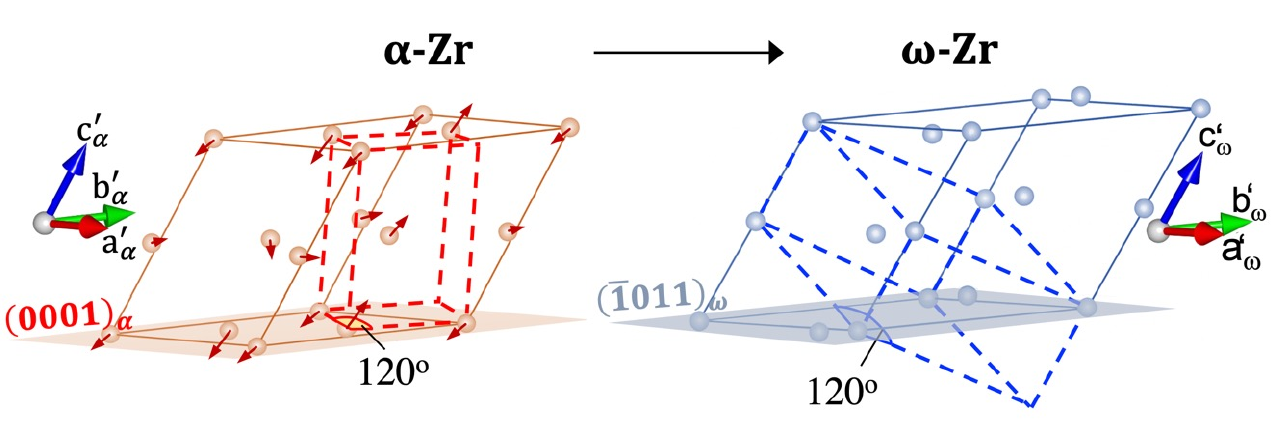}
\caption{The TAO-I mechanism (variant I) is a mapping of two triclinic supercells of $\alpha$ and $\omega-$phases. This pathway is the lowest energy $\alpha\rightarrow\omega$ pathway during homogeneous nucleation. A combination of lattice strain and atomic shuffle transforms the alpha supercell to the omega supercell. The atomic shuffle are shown by the red arrows, with the tail of the arrow showing the original location in the $\alpha-$phase, and the head the final location in the $\omega-$phase. The primitive unit cell (dashed line) along with the alpha supercell is shown in the left panel. The $(0001)_{\alpha}$ plane, shown in red, is fully converted to the $(\bar{1}011)_{\omega}$ plane shown in blue.
	\label{fig:TAO}
}
\end{figure}
The TAO-I mechanism is a mapping of two triclinic supercells of the $\alpha$ and $\omega-$phases using a combination of lattice strain and atomic shuffle. This is the lowest energy pathway under hydrostatic pressure conditions if only homogeneous nucleation is considered \cite{Guan2016}. The TAO-I mechanism produces variant I OR between the $\alpha$ and $\omega$ unit cells given by $(0001)_{\alpha} \vert\vert (01\bar{1}1)_{\omega}$, $[11\bar{2}0]_{\alpha} \vert\vert [\bar{1}011]_{\omega}$. The $\alpha-$supercell (primed) is constructed from the hexagonal basis by the following change of basis \cite{osu1070481734}
\begin{gather*}
\begin{pmatrix}
	\mathbf{a}^{\prime}_{\alpha} \\
	\mathbf{b}^{\prime}_{\alpha} \\
	\mathbf{c}^{\prime}_{\alpha}
\end{pmatrix} = \begin{pmatrix}
	\mathbf{a}_{\alpha}\quad
	\mathbf{b}_{\alpha}\quad
	\mathbf{c}_{\alpha}
\end{pmatrix} 
\begin{pmatrix}
	2 & 1 & 1 \\
	1 & 2 & 1 \\
	0 & 0 & 1
\end{pmatrix}.
\end{gather*}
Similarly, the $\omega-$supercell (primed) is constructed from the hexagonal basis (un-primed) by the following change of basis
\begin{gather*}
\begin{pmatrix}
	\mathbf{a}^{\prime}_{\omega} \\
	\mathbf{b}^{\prime}_{\omega} \\
	\mathbf{c}^{\prime}_{\omega}
\end{pmatrix} = \begin{pmatrix}
	\mathbf{a}_{\omega}\quad
	\mathbf{b}_{\omega}\quad
	\mathbf{c}_{\omega}
\end{pmatrix} 
\begin{pmatrix}
	1 & 1 & 0 \\
	0 & 1 & 0 \\
	1 & 1 & 2
\end{pmatrix}.
\end{gather*}
These supercells and location of atoms are shown in Fig.~\ref{fig:TAO}. The magnitude of strains and atomic shuffle for the TAO-I mechanism can be calculated based on the experimentally measured lattice parameters of the two phases. More details about calculating these strains and shuffles can be found in \cite{osu1070481734}.
For shot $93$ (Figs.~\ref{fig:Fig_1}, \ref{fig:Fig_2}a), the principle strain were calculated to be 
\begin{equation*}
\epsilon_{x} = -0.09989 \quad \epsilon_{y} = 0.0189 \quad \epsilon_{z} = 0.0960.
\end{equation*}
\noindent The strain magnitude for the other shots are of comparable magnitude.
The maximum magnitude of the shuffle for an atom in the unstrained $\alpha-$basis is given by \cite{osu1070481734}
\begin{equation*}
\delta_{\textrm{max}} = \frac{\sqrt{a_{\alpha}^{\prime 2} + c_{\alpha}^{\prime 2}}}{12},
\end{equation*}
while the root mean squared shuffle is given by
\begin{equation*}
\delta_{\textrm{RMS}} = \frac{\sqrt{3a_{\alpha}^{\prime 2} + c_{\alpha}^{\prime 2}}}{6}.
\end{equation*}
For the lattice parameters in shot 93 (Figs.~\ref{fig:Fig_1}, \ref{fig:Fig_2}a) of our experiment, $\delta_{\textrm{max}} = 0.6634$ \AA ~ and $\delta_{\textrm{RMS}} = 0.6026$ \AA. The shuffle magnitudes for the other shots are of comparable magnitude.

\section{\label{sec:Silcock} Silcock mechanism (variant II) during $\alpha\rightarrow\omega$ transformation}
\begin{figure}[!ht]
\centering
\includegraphics[width=1\textwidth]{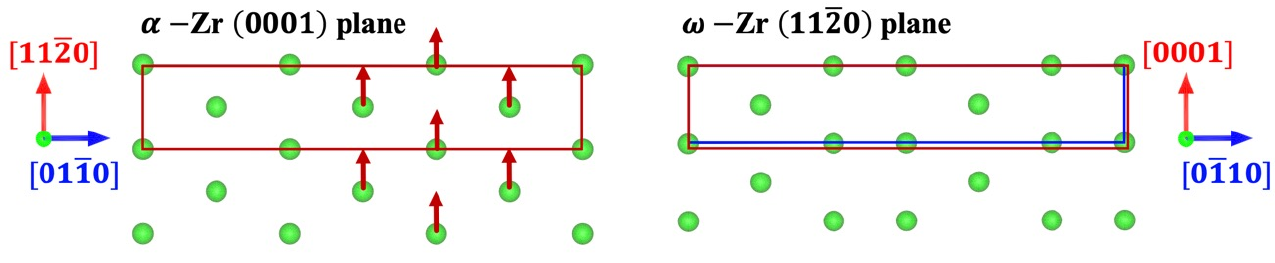}
\caption{The Silcock mechanism (variant II) is a mapping of the a different pair of $\alpha$ and $\omega-$phase supercells. The atomic shuffle is shown by the red arrows, with the tail of the arrow showing the original location in the $\alpha-$phase, and the head the final location in the $\omega-$phase. This is followed by lattice strains mapping the $(0001)_{\alpha}$ plane to the $(11\bar{2}0)_{\omega}$ plane. The $\alpha$ and $\omega-$phase supercells are shown with the solid red and blue line respectively.
	\label{fig:silcock}
}
\end{figure}
The Silcock mechanism is a mapping of two orthorhombic supercells of the $\alpha$ and $\omega-$phase susing a combination of lattice strains and shuffles. This mechanism needs large atomic shuffles but modest lattice strain. The Silcock mechanism produces the variant II OR given by $(0001)_{\alpha} \vert\vert (11\bar{2}0)_{\omega}$, $[11\bar{2}0]_{\alpha} \vert\vert [0001]_{\omega}$. For the Silcock mechanism, the orthorhombic $\alpha-$supercell is constructed from the hexagonal basis by the following change of basis \cite{osu1070481734}
\begin{gather*}
\begin{pmatrix}
	\mathbf{a}^{\prime}_{\alpha} \\
	\mathbf{b}^{\prime}_{\alpha} \\
	\mathbf{c}^{\prime}_{\alpha}
\end{pmatrix} = \begin{pmatrix}
	\mathbf{a}_{\alpha}\quad
	\mathbf{b}_{\alpha}\quad
	\mathbf{c}_{\alpha}
\end{pmatrix} 
\begin{pmatrix}
	1 & 0 & -3 \\
	0 & 0 & -6 \\
	0 & 1 & 0
\end{pmatrix}.
\end{gather*}

Similarly, the orthorhombic $\omega-$supercell is constructed from the hexagonal basis by the following change of basis
\begin{gather*}
\begin{pmatrix}
	\mathbf{a}^{\prime}_{\omega} \\
	\mathbf{b}^{\prime}_{\omega} \\
	\mathbf{c}^{\prime}_{\omega}
\end{pmatrix} = \begin{pmatrix}
	\mathbf{a}_{\omega}\quad
	\mathbf{b}_{\omega}\quad
	\mathbf{c}_{\omega}
\end{pmatrix} 
\begin{pmatrix}
	0 & -1 & 2 \\
	0 & 0 & 4 \\
	-1 & 0 & 0
\end{pmatrix}.
\end{gather*}
These supercells and location of atoms are shown in Fig.~\ref{fig:silcock}.The lattice strains and magnitude of atomic shuffles for the Silcock mechanism can be calculated using the measured lattice parameters of the $\alpha$ and $\omega-$phase. For shot 93 (Figs.~\ref{fig:Fig_1}, \ref{fig:Fig_2}a), the principle strains were calculated to be 
\begin{equation*}
\epsilon_{x} = -0.0669 \quad \epsilon_{y} = -0.0052 \quad \epsilon_{z} = 0.009.
\end{equation*}
The magnitude of shuffle for all atom in the unstrained $\alpha-$basis is equal and is given by \cite{osu1070481734}
\begin{equation*}
\delta_{\textrm{max}} = \delta_{\textrm{RMS}} = \frac{\sqrt{a^{\prime 2}_{\alpha} + c^{\prime 2}_{\alpha}/324}}{4}.
\end{equation*}
For the lattice parameters in shot 93 (Figs.~\ref{fig:Fig_1}, \ref{fig:Fig_2}left) of our experiment, $\delta_{\textrm{max}} = \delta_{\textrm{RMS}} = 0.8243$ \AA.

\section{\label{sec:variantIII} variant III orientation relationship during $\alpha\rightarrow\omega$ transformation}
In addition to variant I and II ORs, we observe the presence of a third, previously experimentally unreported orientation relationship in our measurements. This OR is very closely related to the one reported by Zong et al. \cite{ZONG2014a} in a MD simulation study of shock compression along the \textit{c}-axis of Ti using a modified embedded atom model (MEAM) potential. Our proposed OR, which we refer to as variant III, is $30^{\circ}$ rotated from the one reported in that study and is given by $(0001)_{\alpha} \vert\vert (1\bar{1}00)_{\omega}$, $[11\bar{2}0]_{\alpha} \vert\vert [11\bar{2}3]_{\omega}$. We note that the lattice strains and shuffle leading to this orientation relationship is unknown.

\clearpage
\bibliographystyle{nature}
\bibliography{suppreferences}

\end{document}